\documentclass[twocolumn]{aastex631}

\usepackage{times}
\usepackage{amsmath}
\usepackage{graphicx}
\usepackage{subfigure}
\usepackage{placeins}
\usepackage{hyperref}
\usepackage{gensymb}
\usepackage{upgreek}
\usepackage{soul}

\newdimen\digitwidth    % These five lines change the meaning of
\setbox1=\hbox{0}       % " in the same way, so that it leaves as
\digitwidth=\wd1        % much space as a digit. (Note: All digits
\catcode`"=\active      % are the same width).
\def"{\kern\digitwidth}

\newcommand{\RomanNumeralCaps}[1]
    {\MakeUppercase{\romannumeral #1}}

\newcommand{\MHI}{{\mbox{$M_{\rm H\, \textsc{i}}$}}}

\newcommand{\msun}{$M_{\mathrm{\odot}}$}
\newcommand{\HISS}{H${\scriptscriptstyle\,\mathrm{I}}$}
\newdimen\digitwidth    % These five lines change the meaning of
\setbox1=\hbox{0}       % " in the same way, so that it leaves as
\digitwidth=\wd1        % much space as a digit. (Note: All digits
\catcode`"=\active      % are the same width).
\def"{\kern\digitwidth}

\def\kms{km~s$^{-1}$}
\def\msun{$M_{\odot}$}
\def\HI{H\,{\textsc{\romannumeral 1}}}
\def\RmOne{\RomanNumeralCaps 1}
\def\RmTwo{\RomanNumeralCaps 2}

\def\Nancay{Nan$\c{c}$ay}
\received{\today}
\revised{xxxx, 2021}
\accepted{xxxx}
\submitjournal{ApJ}

\shorttitle{\HI\ Profiles of Massive Galaxy Mergers}
\shortauthors{Zuo et al.}
\graphicspath{{./}{figures/}}

\begin{document}

\title{Massive Galaxy Mergers Have Distinctive Global {H\,\footnotesize{I}} Profiles}

\correspondingauthor{Jinyi Shangguan}
\email{peizuo@pku.edu.cn, shangguan@mpe.mpg.de}

\author[0000-0003-3948-9192]{Pei Zuo}
\affiliation{Kavli Institute for Astronomy and Astrophysics, 
Peking University, Beijing 100871, China}

\affiliation{International Centre for Radio Astronomy Research (ICRAR), University of Western Australia, 35 Stirling Highway, Crawley, WA 6009, Australia}

\author[0000-0001-6947-5846]{Luis C.\ Ho}
\affiliation{Kavli Institute for Astronomy and Astrophysics, 
Peking University, Beijing 100871, China}
\affiliation{Department of Astronomy, School of Physics, Peking University, Beijing 100871, China}

\author[0000-0002-6593-8820]{Jing Wang}
\affiliation{Kavli Institute for Astronomy and Astrophysics, 
Peking University, Beijing 100871, China}
\affiliation{Department of Astronomy, School of Physics, Peking University, Beijing 100871, China}

\author[0000-0002-9066-370X]{Niankun Yu}
\affiliation{Kavli Institute for Astronomy and Astrophysics, 
Peking University, Beijing 100871, China}
\affiliation{Department of Astronomy, School of Physics, Peking University, Beijing 100871, China}

\author[0000-0002-4569-9009]{Jinyi Shangguan}
\affiliation{Max Planck Institute for Extraterrestrial Physics (MPE), Giessenbachstr.1, 85748 Garching, Germany}

\begin{abstract}
The global 21~cm \HI\ emission-line profile of a galaxy encodes valuable information on the spatial distribution and kinematics of the neutral atomic gas.  Galaxy interactions significantly influence the \HI\ disk and imprint observable features on the integrated \HI\ line profile. In this work, we study the neutral atomic gas properties of galaxy mergers selected from the Great Observatories All-sky LIRG Survey.  The \HI\ spectra come from new observations with the Five-hundred-meter Aperture Spherical Telescope and from a collection of archival data.  We quantify the \HI\ profile of the mergers with a newly developed method that uses the curve-of-growth of the line profile.  Using a control sample of non-merger galaxies carefully selected to match the stellar mass of the merger sample,  we show that mergers have a larger proportion of single-peaked \HI\ profiles, as well as a greater tendency for the \HI\ central velocity to deviate from the systemic optical velocity of the galaxy.  By contrast, the \HI\ profiles of mergers are not significantly more asymmetric than those of non-mergers.  

\end{abstract}

\section{Introduction} 
\label{sec:intro}

Galaxy mergers constitute an essential stage of galaxy evolution in a hierarchical universe. \cite{Toomre1972ApJ...178..623T} proposed that mergers transform spirals into ellipticals. Galaxy-galaxy interactions promote gas dissipation, starburst activity, black hole growth, structural assembly, and morphological transformation \citep{Barnes1992ARA&A..30..705B,Kormendy2013ARA&A..51..511K,Mundy2017MNRAS.470.3507M}.  Quantifying the incidence of mergers, the type of mergers (e.g., gas richness and mass ratio), and the dependences on mass and environment are fundamental for the study of galaxy evolution \citep{Darg2010MNRAS.401.1552D, Lotz2010MNRAS.404..575L, Lotz2011ApJ...742..103L, Kampczyk2013ApJ...762...43K}. Multi-wavelength observations of diverse stages and types of mergers demonstrate that most interacting galaxies experience substantial dynamical evolution \citep{Schweizer1986Sci...231..227S, Schweizer1996AJ....111..109S,Hibbard1996AJ....111..655H, Surace2004AJ....127.3235S,Oh2016ApJ...832...69O, Brown2019ApJ...879...17B}, which imparts significant irregularities in the distribution and kinematics of the gas and stars \citep{Barrera-Ballesteros2015A&A...582A..21B,Bloom2017MNRAS.465..123B,Bloom2018MNRAS.476.2339B,Feng2020ApJ...892L..20F}. Numerical simulations \citep{Barnes1996ApJ...471..115B,Barnes2002MNRAS.333..481B, Kapferer2008MNRAS.389.1405K} show that gravitational interactions lead to asymmetric gas distributions and velocity fields in galaxies. Asymmetric features can be long-lasting in the outer regions of the galaxy.  Gas can be stripped by a companion, or it can flow inward due to the torque exerted by the companion or asymmetric structures formed during the interaction.  The complex exchange of gas in different phases can boost the reservoir of molecular gas and replenish the atomic gas by cooling \citep{Moreno2019MNRAS.485.1320M}.

The gas in the interstellar medium of interacting galaxies has been investigated both through observations and theory.  The atomic and molecular gas of merging galaxies can be restructured,
(e.g., \citealt{Horellou2001A&A...376..837H, Yun2001ApJ...550..104Y, Cullen2007MNRAS.376...98C}), and the 
overall gas content can be enhanced (e.g., \citealt{Larson2016ApJ...825..128L, Ellison2018MNRAS.478.3447E}).   Neutral atomic hydrogen (\HI) traces the extended and diffuse gas distribution in galaxies.  The \HI\ disk can be easily distorted by tidal effects, making \HI\ a sensitive tracer of galaxy interactions (e.g., \citealt{Yun1994Natur.372..530Y, Kornreich2000AJ....120..139K, Reichard2008ApJ...677..186R}).  Images show that the \HI\ of merging galaxies produce bridges and tails that can extend to several hundred kpc (e.g., \citealt{Horellou2001A&A...376..837H, Yun2001ApJ...550..104Y, Cullen2007MNRAS.376...98C}), as well as forming intra-group clouds that can account for $\sim$10\% of the \HI\ in the entire interacting group \citep{Serra2015MNRAS.452.2680S}.  Interferometric observations can sensitively probe perturbations on the \HI\ (\citealt{Wang2013MNRAS.433..270W}), but they are not widely available. While less information is contained in single-dish \HI\ spectra, they nevertheless provide the atomic gas content, radial velocity, line width, profile asymmetry, and concentration \citep{Yu2020ApJ...898..102Y}.  Single-dish surveys such as the \HI\ Parkes All-Sky Survey (HIPASS: \citealt{Koribalski2004AJ....128...16K}), xGASS \citep{Catinella2012MNRAS.420.1959C, Catinella2018MNRAS.476..875C}, and ALFALFA \citep{Haynes2011AJ....142..170H, Haynes2018ApJ...861...49H} have published tens of thousands of spectra for nearby galaxies. 

The exact influences of galactic mergers on the shape of \HI\ profiles have been inconclusive. In non-merger galaxies, gas accretion from the large-scale environment, minor mergers of satellite galaxies, encounters with a flyby galaxy, and interactions with the dark matter halo are all possible mechanisms that can perturb the distribution of \HI\ and induce profile asymmetry \citep{Mapelli2008MNRAS,Sancisi2008AARv,Lagos2018MNRAS}.  In denser environments, the mechanisms that enhance \HI\ asymmetry include ram pressure stripping \citep[e.g.,][]{Gunn1972ApJ,Kenney2004AJ}, harrasment \citep[e.g.,][]{Moore1996Natur}, tidal stripping (e.g., \citealt{Moore1999MNRAS,Koribalski2009MNRAS.400.1749K,English2010AJ....139..102E}), and galactic mergers (e.g., \citealt{Bok2019MNRAS.484..582B}).  

Previous investigations have not yielded a clear relation between \HI\ profile asymmetry and the dynamical state of the galaxy. For example, even relatively isolated galaxies can exhibit high levels of profile asymmetry \citep{Espada2011A&A...532A.117E}, while interacting systems can have symmetric profiles \citep{Watts2021MNRAS.504.1989W}. \citet{Reynolds2020MNRAS.499.3233R} found no significant correlation between profile asymmetry and environment in their study of HIPASS galaxies, although within the xGASS sample profile asymmetry seems to be related to whether a galaxy is a central or a satellite of a group \citep{Watts2020MNRAS.499.5205W}.  Within the local volume, profile asymmetry appears to increase with the local galactic density \citep{Reynolds2020MNRAS.493.5089R}.  In their comparative analysis of galaxy pairs and isolated galaxies, \citet{Bok2019MNRAS.484..582B} found that merger activity indeed enhances asymmetry in the global \HI\ profile, but in order to avoid observational confusion, they excluded very close pairs, and thus it is unclear whether the results are biased against close encounters, particularly late-stage mergers.  

Recently, \citet{Yu2020ApJ...898..102Y} developed a new method to analyze one-dimensional spectral profiles, borrowing from the concept of the curve-of-growth (CoG), which is widely used in two-dimensional image analysis. In this paper, we apply the CoG method to analyze the integrated \HI\ spectra of a subset of merger galaxies selected from the Great Observatories All-sky LIRG Survey (GOALS; \citealt{Armus2009PASP..121..559A}), a well-defined, infrared-selected sample consisting of 202 luminous, massive, low-redshift ($z \lesssim 0.1$), galaxies that are primarily mergers in different evolutionary stages.  GOALS is arguably the best-studied sample of nearby starburst galaxies, one that possesses a large amount of ancillary multi-wavelength data (e.g., Spitzer: \citealt{Inami2013ApJ...777..156I}; Herschel: \citealt{Chu2017ApJS..229...25C}; HST: \citealt{Haan2011AJ....141..100H}; ALMA: \citealt{Xu2015ApJ...799...11X}; Chandra: \citealt{Iwasawa2011A&A...529A.106I}).  The molecular gas \citep{Larson2016ApJ...825..128L,Yamashita2017ApJ...844...96Y} and dust \citep{Herrero-Illana2019A&A...628A..71H,Shangguan2019ApJ...870..104S} properties have been investigated in detail.  Missing, to date, is a systematic study of the \HI\ properties.  

We will statistically compare the \HI\ spectra of mergers with those of a control sample of non-mergers, with the aim of ascertaining whether the \HI\ profile can provide any useful tracers that can distinguish the two populations.  In Section~\ref{sec:data}, we define the sample and introduce the new FAST observations.  The CoG method is briefly introduced in Section~\ref{sec:CoG}.  We compare the \HI\ properties of mergers and control galaxies in Section~\ref{sec:4}.  A summary of the main results is provided in Section~\ref{sec:summary}. This work adopts the following parameters for a $\Lambda$CDM cosmology: $\Omega_m = 0.308$, $\Omega_\Lambda = 0.692$, and $H_{0}=67.8$ km s$^{-1}$ Mpc$^{-1}$ \citep{Planck2016}.

\section{Data} 
\label{sec:data}

We analyze 56 mergers contained in the GOALS sample, seven based on new \HI\ observations obtained with FAST and 49 from published \HI\ spectra collected from the NASA/IPAC Extragalactic Database (NED)\footnote{\url{http://ned.ipac.caltech.edu/forms/SearchSpectra.html}}.  We adopt the morphological classifications, based on optical and near-infrared images, from the studies of \cite{Stierwalt2013ApJS..206....1S}, \cite{Larson2016ApJ...825..128L}, and \cite{Jin2019ApJS..244...33J}.  In addition, we analyze a control sample of non-mergers drawn from a representative sample of nearby galaxis recently studied by \citet{Yu2020ApJ...898..102Y}.  As the non-mergers have preferentially lower stellar masses than the mergers, we restrict the comparison only to the 80 control galaxies that statistically match the merger sample over the stellar mass range $10^{10.25}-10^{11.25}\,M_\odot$ (Table \ref{tbl:paras-list_1}).  

\subsection{FAST Observations} 
\label{subsec:obs_F}

New FAST data for eight sources (Table~\ref{tbl:fast-list}; Figure~\ref{fig:spec_FAST}) from the GOALS sample were acquired during a pilot cycle of ``shared-risk'' observations (PI: Ho; project code: 2019a-017-S).  Six of the targets did not have prior \HI\ observations, while two were deliberately chosen to be redundant with published observations in order to verify the performance of FAST. One source, UGC 1845, turns out to be a non-merger galaxy, and we did not include it in the following analysis. The observations, carried out using the 19-beam receiver at $L$ band ($1.05-1.45$~GHz) in August 2019, were taken with the entire wide (500~MHz) band using the on-off mode and the central (M01) beam, using 300~s for each on and off pointing and $\sim 30$~s for switching between them.  The beam size is $2\farcm 9 \times 2\farcm 9$.  Using $\sim 1$~s sampling, the dual-polarization observations provide 0.475~kHz ($\sim 0.1$~\kms) spectral resolution covering 1,048,576 channels.  We calibrate the flux intensity using data from the on-off noise diode, which were acquired during the observations, and we apply aperture efficiency correction based on \cite{Jiang2019SCPMA..6259502J}.  Scans and channels with radio frequency interference (RFI) were flagged.  The data for each polarization were accumulated and averaged.  A polynomial of order $1-3$ was used to fit the spectrum and subtract the baseline.  The spectra were binned to a velocity resolution of $\sim 6.4$~\kms\ in order to be consistent with the spectral resolution of the archival data.  The analysis results of the new \HI\ data are listed in Table~\ref{tbl:paras-list}. 

Two of the targets, IRAS~17578$-$0400 and IRAS~F17138$-$1017, have prior observations by the \Nancay\ telescope, affording an opportunity for a direct comparison with the FAST data (Figure~\ref{fig:spec_comp}).  The FAST spectra show $\sim 35\%$ and $\sim 31\%$ higher total \HI\ flux than those from \Nancay, for IRAS~17578$-$0400 and IRAS~F17138$-$1017, although the line profiles and velocity ranges are consistent.  While typical uncertainties in total flux are expected to be 10\%--15\%, accounting for calibration ($\sim$6\%, \citealt{Liu2021}) effects such as beam attenuation, pointing, flux calibration, and \HI\ self-absorption \citep{Springob2005ApJS..160..149S}, it is likely that the early \Nancay\ measurements underestimate fluxes for the brightest, more extended sources \citep{Haynes2011AJ....142..170H}.  \citet{van2016A&A...595A.118V} found that the multi-beam observations of Arecibo and Parkes yield fluxes $\sim 35\%-45\%$ higher than those of the single-receiver observations of \Nancay.  This excess is similar to those we find above.  Nevertheless, there is good consistency for the asymmetry and concentration parameters of the \HI\ profiles from the FAST and \Nancay\ spectra.

%edited by LCH 2021.06.17
%edited by LCH 2021.11.29

% \FloatBarrier
\begin{figure*}[ht!]
\setcounter{figure}{0}
\centering
\epsscale{0.45}
\plotone{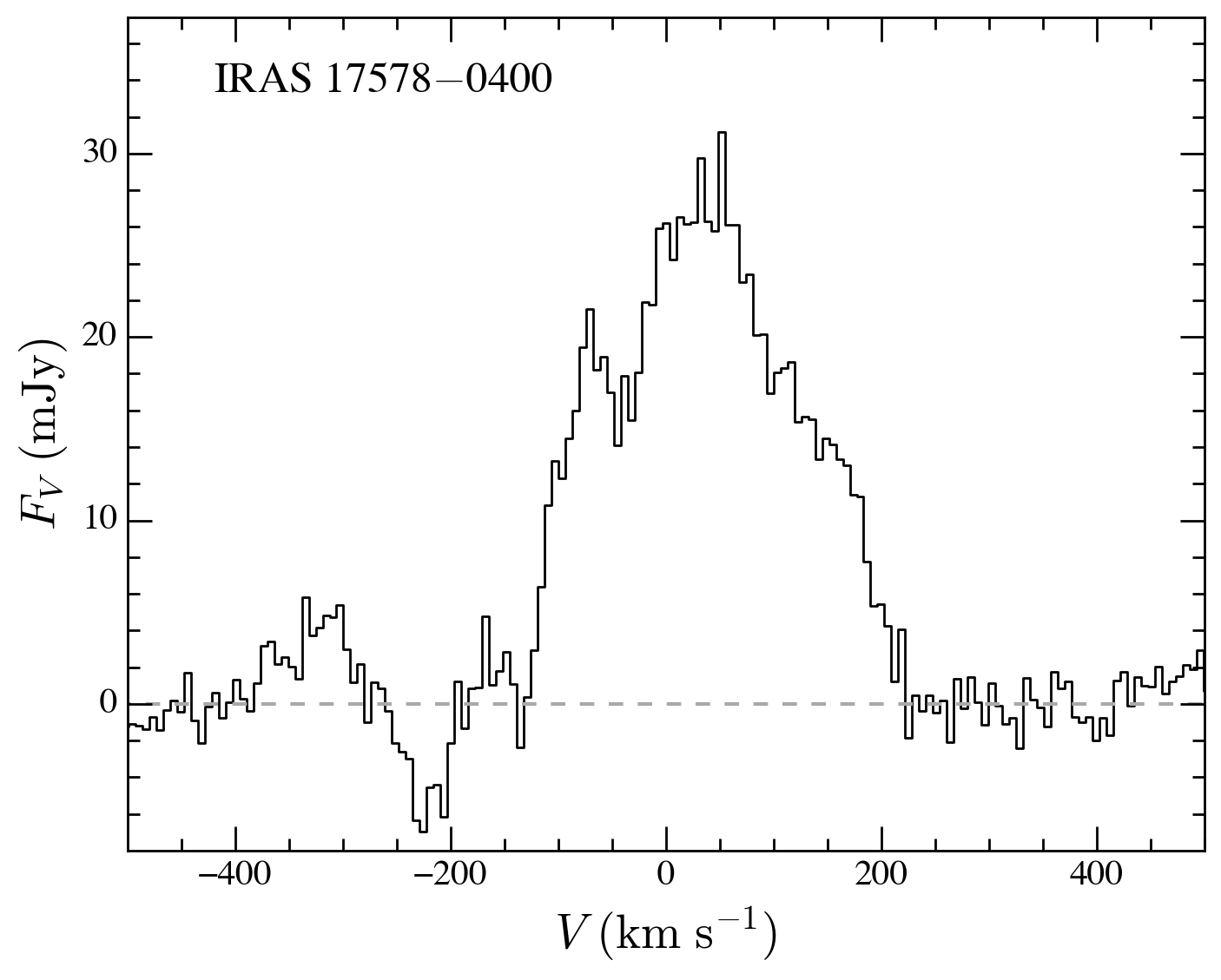}
\epsscale{0.45}
\plotone{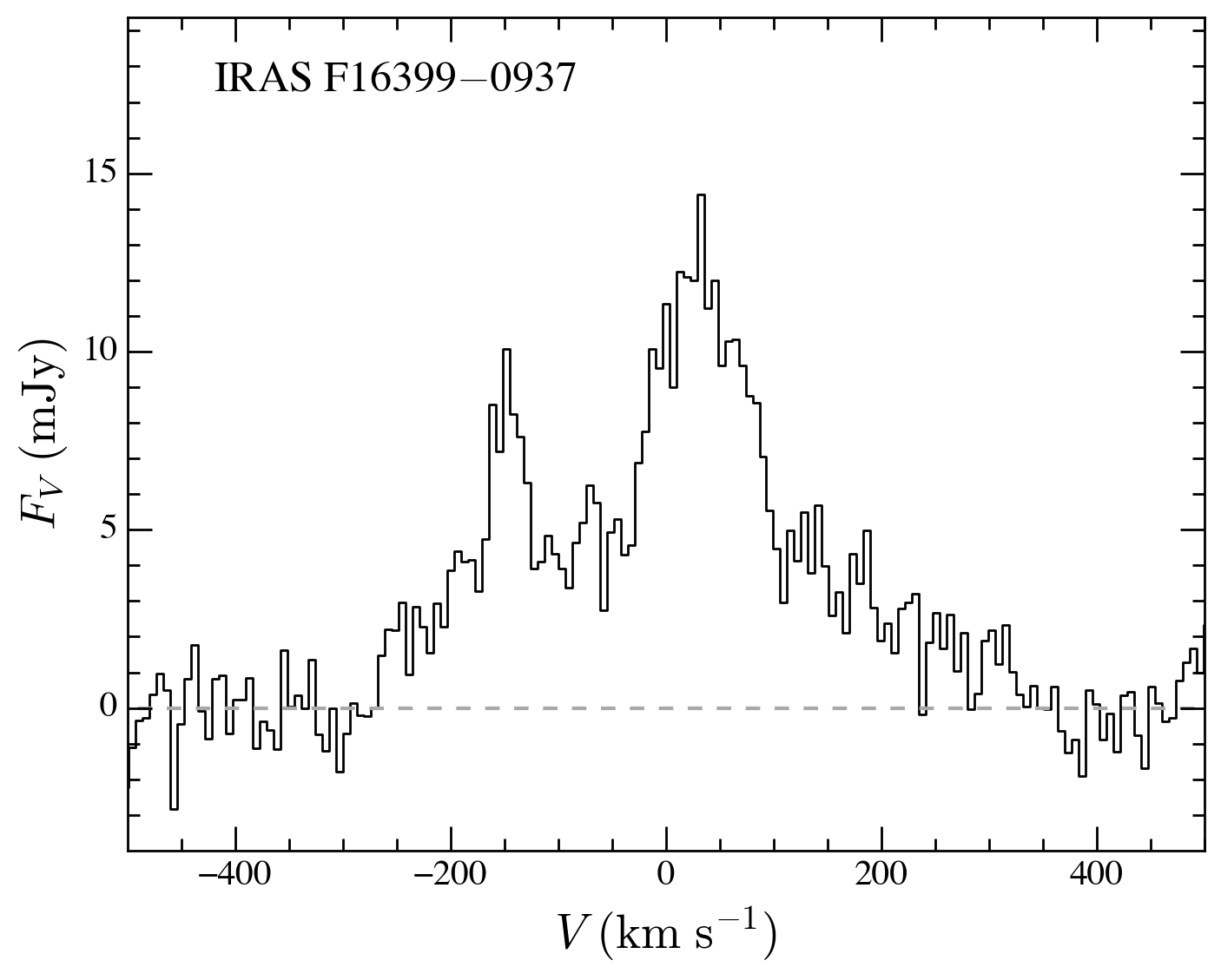}

\epsscale{0.45} %**
\plotone{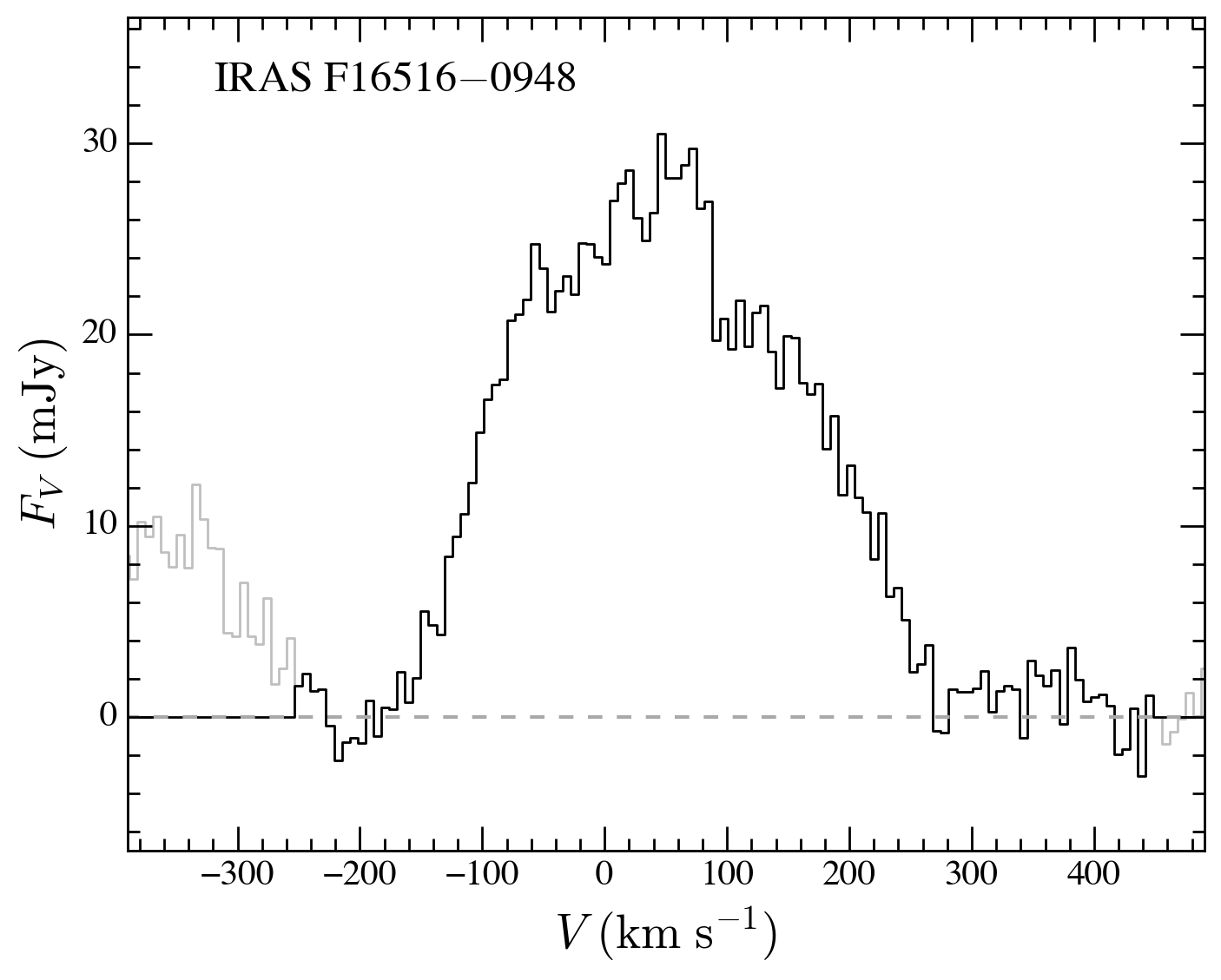}
\epsscale{0.45}
\plotone{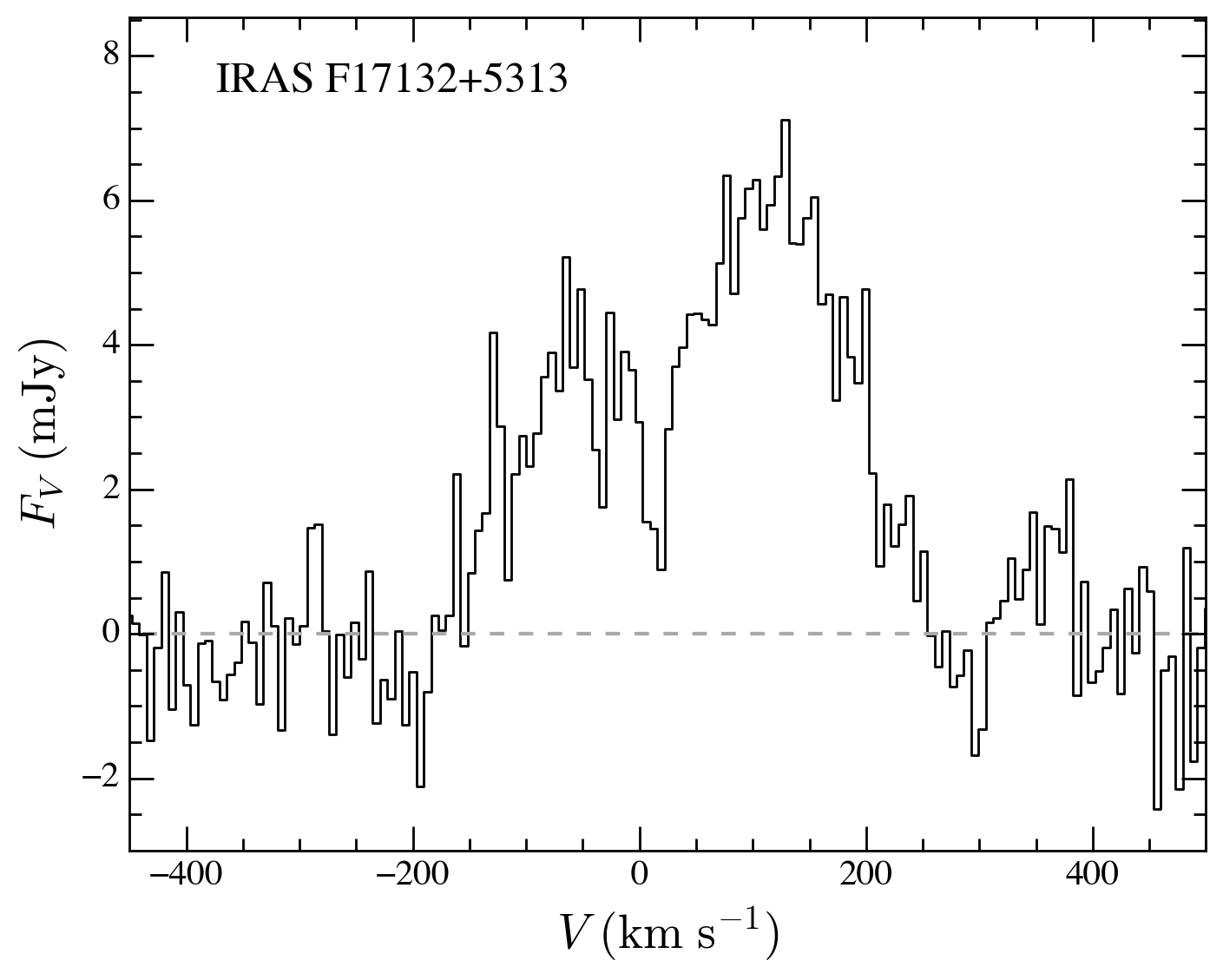}

\epsscale{0.45}
\plotone{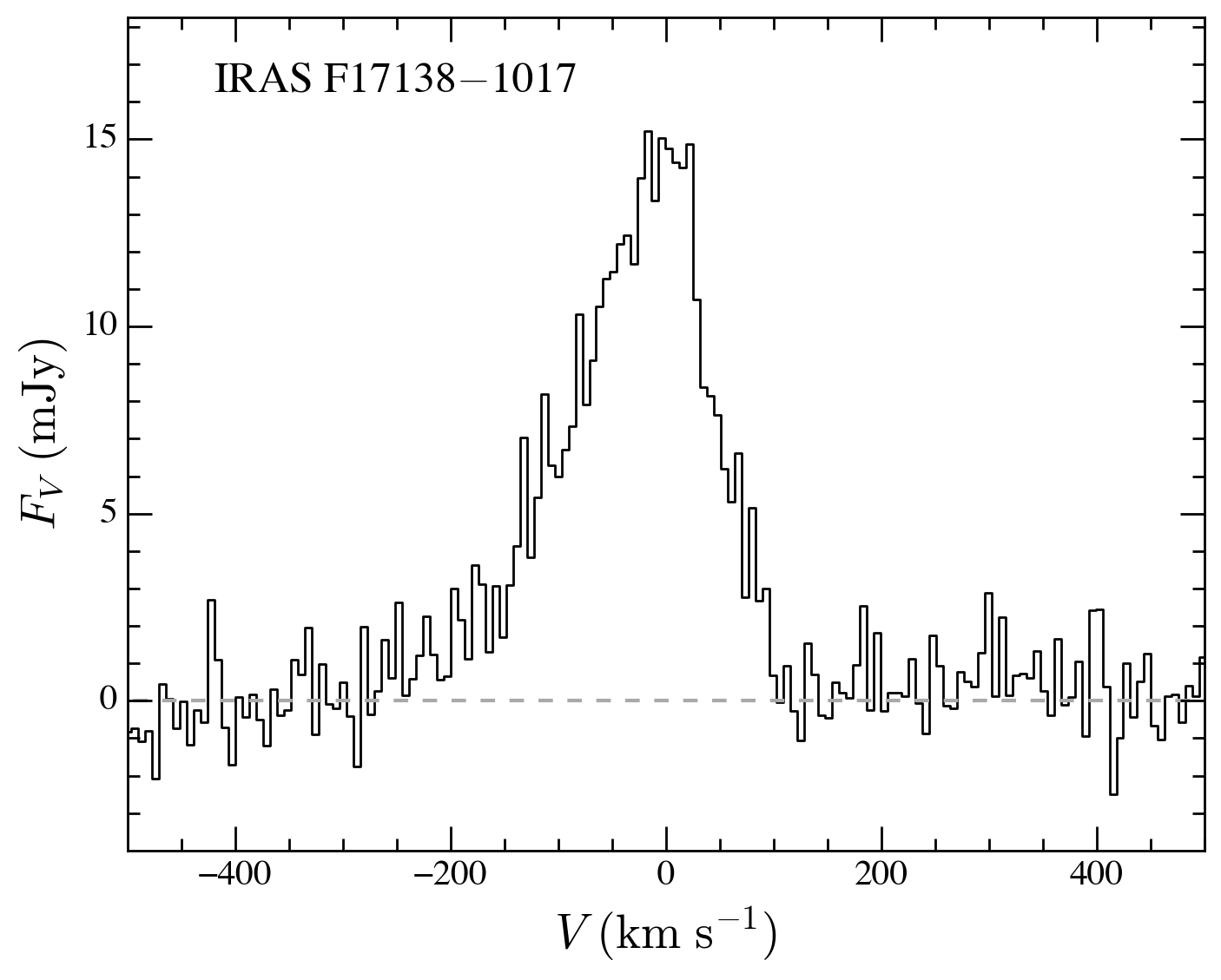}
\epsscale{0.45}
\plotone{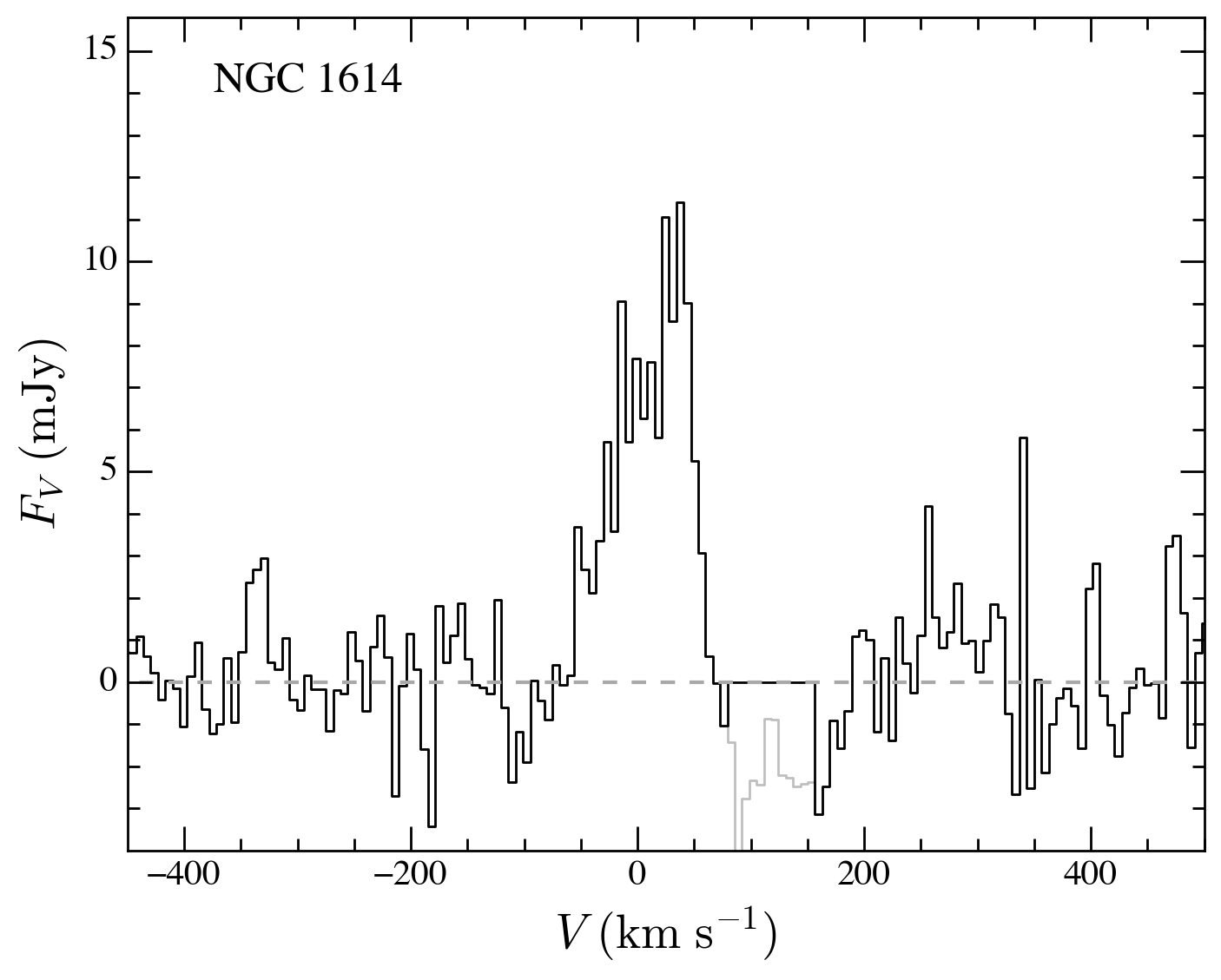}

\epsscale{0.45}
\plotone{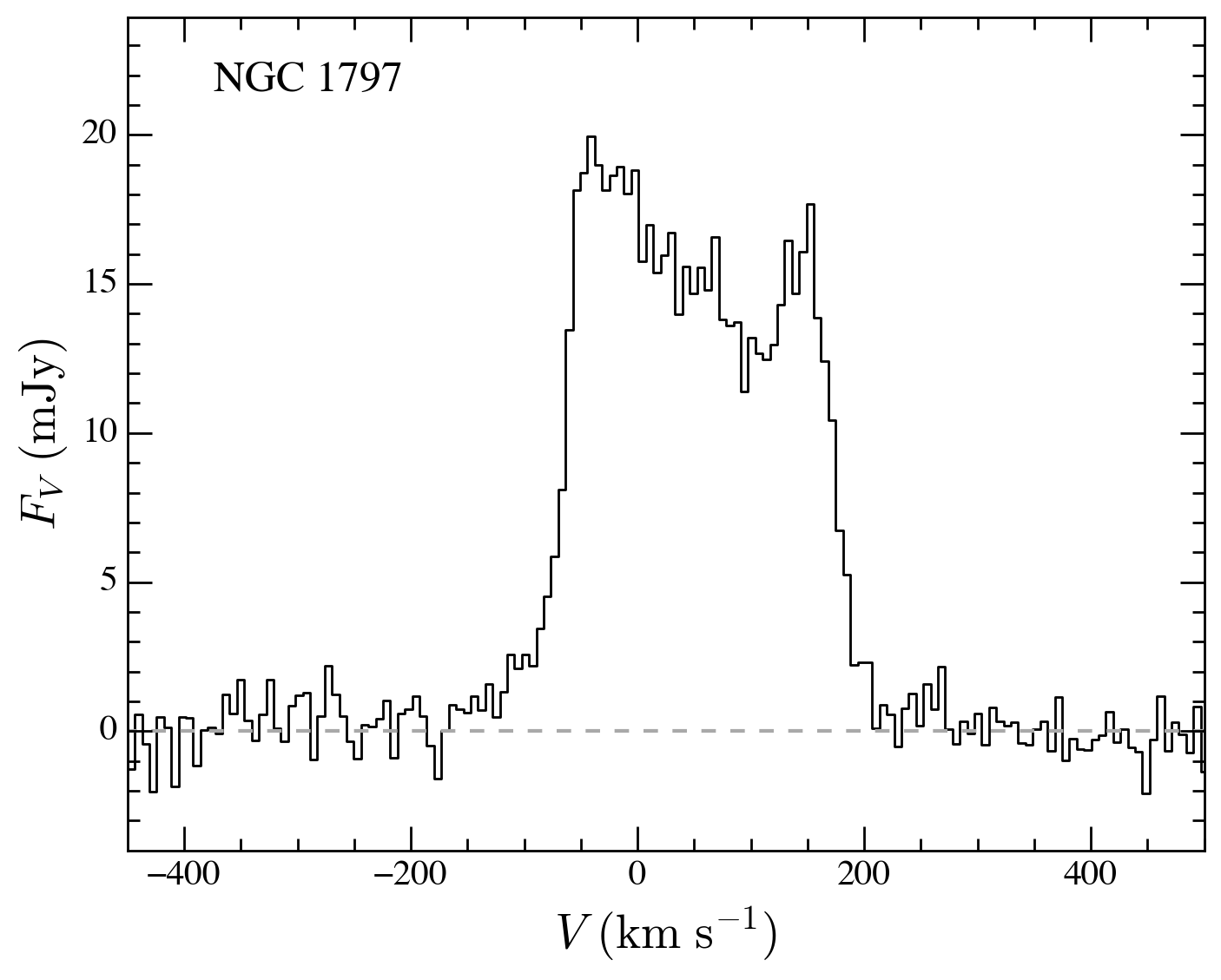}
\epsscale{0.45}
\plotone{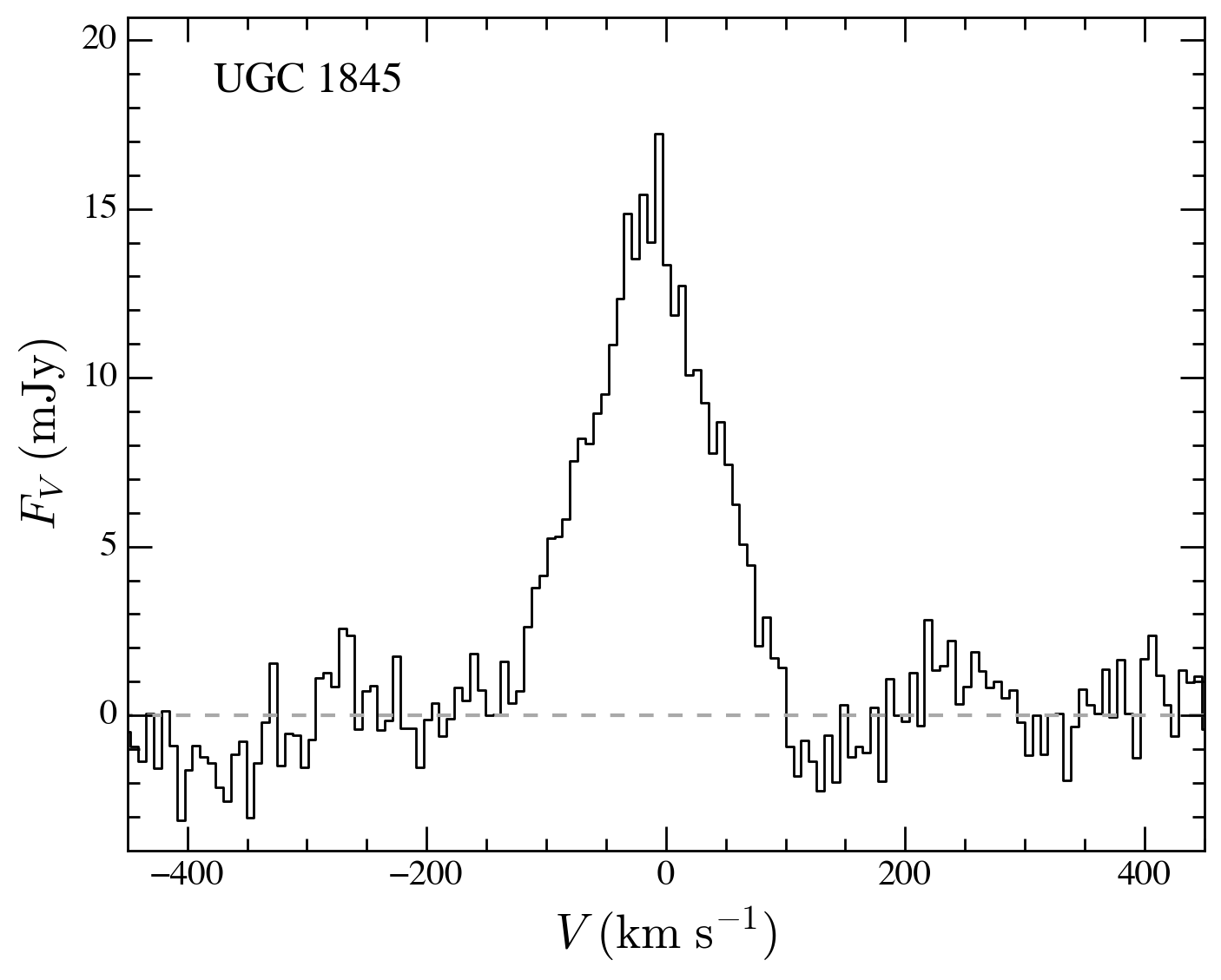}

%YY: UGC 1845 is a non-merger galaxy in the GOALS sample. 
\caption{New FAST \HI\ spectra of eight galaxies from the GOALS sample. 
The velocity is shifted to the rest-frame according to the optical redshift of the target, $V=c\,[(f_0 - f) / f - z]$, where $f_0=1.420~\mathrm{GHz}$ is the rest frequency of \HI\ line, $f$ is the observed frequency, $z$ is the redshift from Table \ref{tbl:fast-list}, and $c$ is the speed of the light. The channels contaminated by RFI are in grey.
}
% 5. Figure 1: Thank you for adding the explanatory text to the figure caption. However, the spectra are not shifted to the systemic velocity, but rather the rest-frame (i.e. v~0 km/s). Please also mention that you are using the 'optical z from Table 1' so that it's clear where the redshift used in the quoted formula comes from.

\label{fig:spec_FAST}
\end{figure*}

\label{fig:spec_FAST}
%edited by LCH 2021.06.08
%edited by LCH 2021.11.29

\begin{deluxetable*}{crrccrrccc}[ht!]%D@{$\pm$}DrD@{$\pm$}Drrrcl}
\tablenum{1}
\centering
\small\addtolength{\tabcolsep}{-1.5pt}
\tablecolumns{11}
\tablecaption{New FAST \HI\ Observations} %ZA  integration time cycle 
\tablehead{
\colhead{Galaxy} & 
\colhead{R.~A.} & 
\colhead{Decl.} & 
\colhead{$z$} & 
\colhead{$D_L$} & 
\colhead{log $M_*$} & 
\colhead{$\nu_c$} &
\colhead{Quality} & \\
%\colhead{Zenith Angle}
\colhead{} &
\colhead{(J2000)} & 
\colhead{(J2000)} &
\colhead{} &
\colhead{(Mpc)} &
\colhead{(\msun)} &
\colhead{(MHz)} &
\colhead{} & 
\colhead{} \\
\colhead{(1)} &
\colhead{(2)} &
\colhead{(3)} &
\colhead{(4)} &
\colhead{(5)} &
\colhead{(6)} &
\colhead{(7)} &
\colhead{(8)} &
}
\decimals
\startdata  % 8-1-2=5 objects
IRAS~17578$-$0400\tablenotemark{a} & 18:00:28.61 & $-$04:01:16.3 & 0.0140 & 70.6 & 10.27 $\pm$ 0.20 & 1400.79 & \RmOne \\
IRAS~F16399$-$0937 & 16:42:40.11 & $-$09:43:13.7 & 0.0270 & 131.5 & 11.00 $\pm$ 0.20 & 1383.06 & \RmTwo \\
IRAS~F16516$-$0948 & 16:54:23.72 & $-$09:53:20.9 & 0.0227 & 111.6 & 10.73 $\pm$ 0.20 & 1388.88 & \RmTwo \\
IRAS~F17132$+$5313 & 17:14:20.45 & $+$53:10:31.6 & 0.0509 & 238.9 & 11.00 $\pm$ 0.20 & 1351.61 & \RmTwo \\
IRAS~F17138$-$1017\tablenotemark{a}  & 17:16:35.68 & $-$10:20:40.5 & 0.0173 & 86.6 & 10.72 $\pm$ 0.20 & 1396.25 & \nodata \\
NGC~1614 & 04:33:59.95 & $-$08:34:47.0 & 0.0159 & 69.9 & 10.85 $\pm$ 0.20 & 1398.17 & \RmOne, \RmTwo \\
NGC~1797 & 05:07:44.84 & $-$08:01:08.7 & 0.0149 & 65.7 & 10.61 $\pm$ 0.20 & 1399.55 & \nodata \\
UGC~1845\tablenotemark{b} & 02:24:07.97 & $+$47:58:11.9 & 0.0156 & 69.1 & 10.57 $\pm$ 0.20 & 1398.59 & \RmOne, \RmTwo \\
%UGC~1281 & 156 & 2 & 38.98 & 6.16 & 55 & 10 & 51 & 3 & 46 & 2 & 40 & 2 & 1.03 & 0.07 & 1.06 & 0.11 & 3.44 & 0.34 & 578.0 & 8.44 & 0.13 & 9.63 & 0.14 & 1, 2 \\
%UGC~2023 & 602 & 2 & 17.23 & 2.73 & 64 & 76 & 56 & 55 & 50 & 45 & 44 & 39 & 1.00 & 0.07 & 1.03 & 0.10 & 4.31 & 0.43 & 351.9 & 8.62 & 0.13 & 9.77 & 0.87 & 2 \\
%%NGC 0838 & 02:09:38.53 & $-$10:08:48.1 & 0.0128 \\
%NGC 1797 & 05:07:44.84 & $-$08:01:08.7 & 0.0149 \\
\enddata

\tablecomments{
Column (1): Galaxy name. 
Columns (2)--(3) Equatorial coordinates (J2000).
Column (4): Redshift from NED. 
Column (5): Luminosity distance from \citet{Shangguan2019ApJ...870..104S}, derived from the heliocentric velocity corrected for the three-attractor flow model of \citet{Mould2000ApJ...529..786M} using $\Omega_m=0.308$, $\Omega_{\Lambda}=0.692$, and $H_0=67.8$ \kms\ Mpc$^{-1}$ \citep{Planck2016}.  
Column (6): Stellar mass from \citet{Shangguan2019ApJ...870..104S}. 
Column (7): Central frequency for the observation. 
Column (8): Quality flag: \RmOne\ = scans with strong radio frequency interferences removed; \RmTwo\ = channels with strong radio frequency interferences flagged. 
\tablenotetext{a}{Has previous H~I observation.}
\tablenotetext{b}{This source is a non-merger galaxy in the GOALS sample, and we do not include it in the analysis. }
}
\label{tbl:fast-list}
\end{deluxetable*}

\begin{figure*}[ht!]
\epsscale{0.6}
\plotone{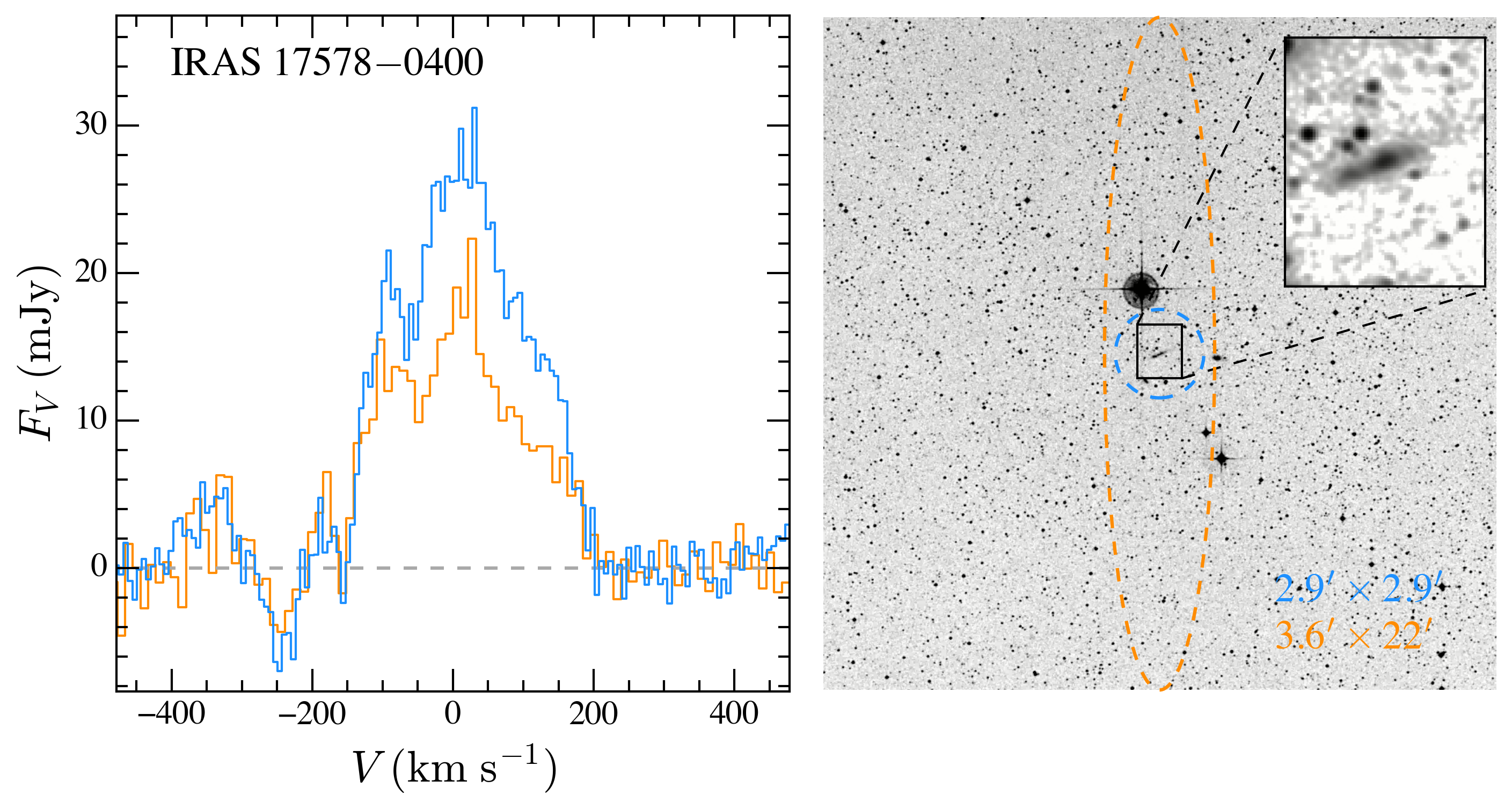}
\epsscale{0.6}
\plotone{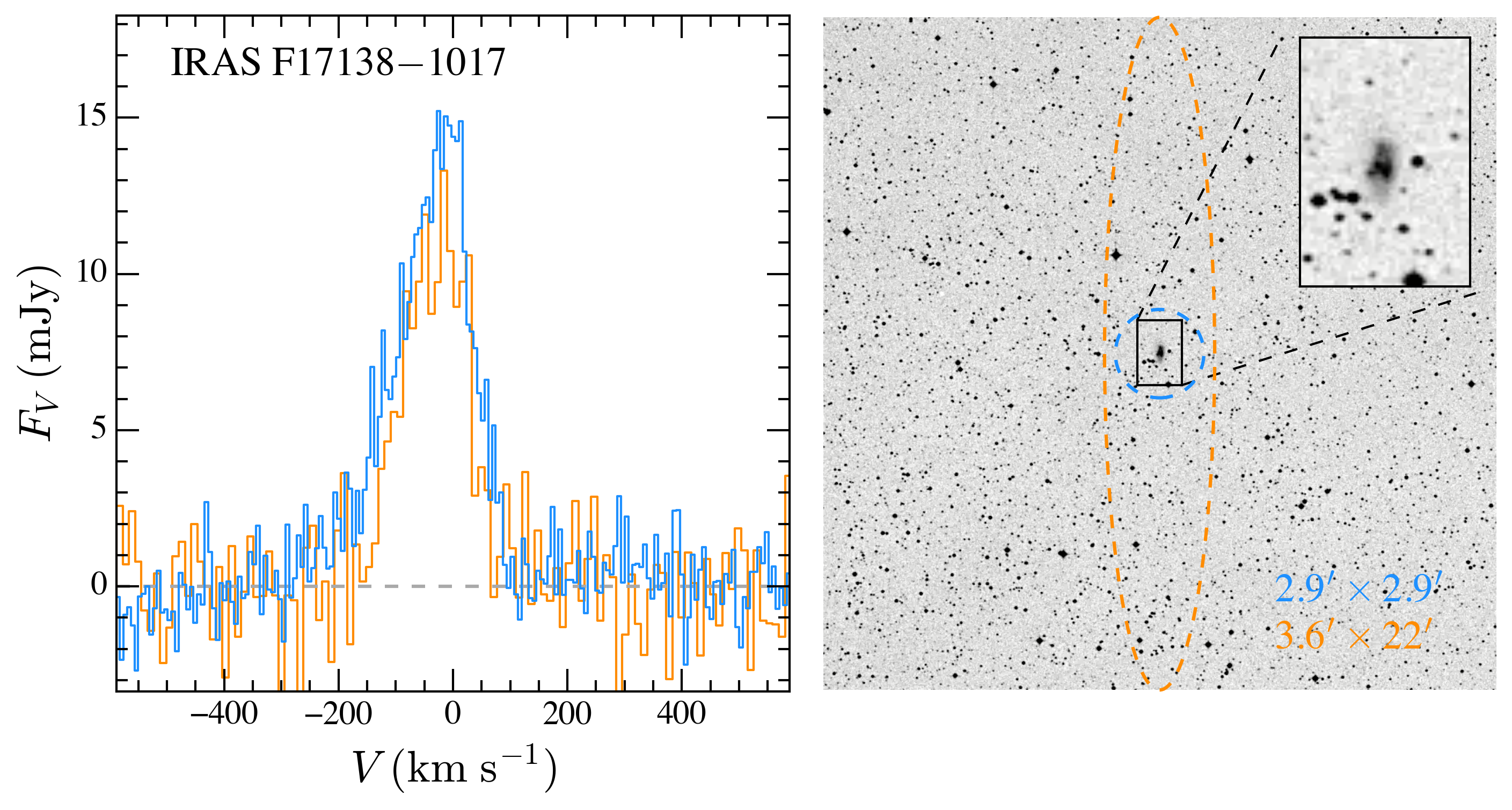}
\caption{Comparison of \HI\ profiles from the literature (orange) with new FAST observations (blue). The right panel for each object displays the optical (4680~\r{A}) Digitized Sky Survey image from NED, with the zoom-in highlighting the morphology of the source in the $J$ band \citep[2MASS;][]{Skrutskie2006AJ....131.1163S}.  The beam size is shown as an orange ellipse for the \Nancay\ observations and as a blue circle for the FAST observations.
}
\label{fig:spec_comp}
\end{figure*}

\subsection{Data from the Literature} 
\label{subsec:obs_achv}

With the aid of NED, we located single-dish \HI\ spectra for 82 GOALS sources.  After excluding 18 sources classified by  \citet{Stierwalt2013ApJS..206....1S}, \citet{Larson2016ApJ...825..128L}, and \citet{Jin2019ApJS..244...33J} as non-mergers and another 13 that exhibit prominent \HI\ absorption, which is frequently associated with active galactic nuclei with strong radio continuum emission \citep{Gereb2015A&A...575A..44G, Glowacki2017MNRAS.467.2766G,Maccagni2017A&A...604A..43M} and radio-loud galaxy mergers \citep{Dutta2019MNRAS.489.1099D}, we are left with 51 mergers (Table~\ref{tbl:achv-list}).  Among these, two have new FAST observation (Section~\ref{subsec:obs_F}; Figure~\ref{fig:spec_comp}), and we adopt the new data in the subsequent analysis.  

If multiple spectra of the same target are available from the same telescope, we chose the one with the highest velocity resolution and sensitivity.  When more than one \HI\ spectrum exists, we adopted the one observed with the smallest beam that can cover the entire galaxy merger system. Meanwhile, we confirm through visual inspection of optical images that there is no contamination from nearby sources within the \HI\ beam and within $\sim$1000 \kms\ of \HI\ central velocity. Most of the spectra were obtained using Arecibo, which, with a beam of $\sim 3\farcm5 \times 3\farcm5$, has the highest sensitivity and spatial resolution.  By contrast, Parkes has a $15\farcm5 \times 15\farcm5$ beam, while that of \Nancay\ is $3\farcm 6 \times 22\arcmin$.  The spectral resolution of the final spectra ranges from 6.6 to 22.0~km~s$^{-1}$, with a median value of 10.5~km~s$^{-1}$.  There are 21 archival \HI\ spectra from Arecibo, 14 from Green Bank Telescope (GBT) 90-m, 11 from \Nancay, two from Parkes, two from Jodrell Bank Lovell, and one from the GBT 110-m.

To construct the control sample of non-mergers, we begin with the representative sample of 269 nearby galaxies whose integrated \HI\ profiles were studied by \cite{Yu2020ApJ...898..102Y}.  We carefully inspect optical images to exclude objects with visibly disturbed features and nearby companions projected within the \HI\ beam, as well as sources exhibiting spectral line contamination within $\pm$1000 \kms\ of the central velocity. Seven targets have apparent companions with projected separations of 50--100~kpc, but they are included because the companions are outside of the telescope beam.  We verified that our main conclusions are not affected by this choice.  In view of the expected perturbed peculiarities and elevated star formation rates of the mergers, it is impractical when selecting a comparison sample to control for factors such as surface brightness, star formation rate, and inclination angle (e.g., \citealt{Wang2013MNRAS.433..270W, Wang2020ApJ...890...63W}). Instead, we focus on stellar mass, choosing $M_* = 10^{10.25}-10^{11.25}\,M_\odot$ as the range that best matches the final sample of 45 mergers and the control sample of 80 non-mergers.  The stellar masses of the GOALS galaxies were derived by \cite{Shangguan2019ApJ...870..104S} using 2MASS $J$-band magnitudes and a mass-to-light ratio estimated from the $B-I$ color and the calibrations of \cite{Bell2003ApJS..149..289B}.  We use a similar strategy for the non-mergers, in this instance using 2MASS $K_s$-band magnitudes and assuming $g-i=1.2$ mag, a color typical of the control galaxies estimated using Sloan Digital Sky Survey data \citep{Ahumada2020ApJS..249....3A}.  All stellar masses are referenced to the stellar initial mass function of \cite{Chabrier2003PASP..115..763C}.  We define five stellar mass bins [$\log (M_*/M_\odot) = 10.25-10.45, 10.45-10.60, 10.60-10.80, 10.80-10.95,$ and $10.95-11.25$]. The width of each bin is adjusted to ensure that there are enough galaxies in both the control and merger sample. For each target in the merger sample, we randomly select a galaxy in the same stellar mass bin from the control sample. We use a Monte Carlo approach to resample the merger and control galaxies 1000 times in order to quantify the statistical uncertainty (see Section~\ref{sec:4}). After one-to-one matching in stellar mass, the merger and non-merger samples both contain 43 galaxies. One particular realization of the matched samples is shown in Figure~\ref{fig:Hist_MStar}. Figure~\ref{fig:opt-hi} gives the \HI\ profiles and optical morphologies of example control and merger galaxies. 

%edited by LCH 2021.06.08
%edited by LCH 2021.11.29

% \startlongtable
\begin{deluxetable*}{crrcccccccccc}[ht!]
\tabletypesize{\normalsize}
\tablenum{2}
\centering
\small\addtolength{\tabcolsep}{-2.5pt}
\tablecolumns{11}
\tablecaption{Basic Information and \HI\ Data of Galaxy Mergers from the Literature}
\tablehead{
\colhead{Galaxy} & 
\colhead{R.~A.} & 
\colhead{Decl.} & 
% \colhead{Type} & 
\colhead{$z$} & 
\colhead{$D_L$} & 
\colhead{log $M_*$} & 
\colhead{log $M_\mathrm{\HI}$} & 
\colhead{Beam size} & 
\colhead{$v_\mathrm{inst}$} & 
\colhead{Ref.} \\
\colhead{} & 
\colhead{(J2000)} & 
\colhead{(J2000)} &
% \colhead{} & 
\colhead{} & 
\colhead{(Mpc)} & 
\colhead{($M_{\odot}$)} & 
% \colhead{($L_{\odot}$)} & 
\colhead{($M_{\odot}$)} & 
\colhead{(\arcmin)} & 
\colhead{(\kms)} \\
\colhead{(1)} & 
\colhead{(2)} & 
\colhead{(3)} & 
\colhead{(4)} & 
\colhead{(5)} & 
\colhead{(6)} & 
\colhead{(7)} & 
\colhead{(8)} & 
\colhead{(9)} &
\colhead{(10)} & 
% \colhead{(11)} & 
% \colhead{(12)} 
\\ 
}
\decimals
\startdata  %64 objects
CGCG~043-099 & 13:01:49.9 & $+$04:20:01 & 0.0375 & 180.2 & 10.87 $\pm$ 0.20 & 10.17 & 3.3 & 8.3 & 1 \\ % rG
CGCG~436-030\tablenotemark{a} & 01:20:01.4 & $+$14:21:35 & 0.0312 & 137.9 & 10.59 $\pm$ 0.20 & $>9.57$ & 3.3 & 8.3 & 1 \\
CGCG~448-020\tablenotemark{a} & 20:57:23.3 & $+$17:07:34 & 0.0361 & 165.9 & 10.96 $\pm$ 0.20 & $>$10.03  & 3.3 & 8.3 & 1 \\ 
CGCG~465-012 & 03:54:16.4 & $+$15:55:44 & 0.0222 & 97.3 & 10.67 $\pm$ 0.20 & 9.42 & 3.3 & 8.3 & 1 \\
IC~2810 & 11:25:47.3 & $+$14:40:23 & 0.0342 & 162.8 & 11.19 $\pm$ 0.20 & 9.91 & 3.3 & 8.3 & 1 \\
IC~5298 & 23:16:01.7 & $+$25:33:33 & 0.0274 & 122.7 & 10.87 $\pm$ 0.20 & 9.80  & 3.3 & 8.3 & 1 \\
\enddata
\tablecomments{
Column (1): Galaxy name.  
Columns (2)--(3): Equatorial coordinates (J2000).
% Column (4): galaxy type; the symbol ``*'' denotes non-merger galaxy reclassified considering information of morphology from \citet{Stierwalt2013,Larson2016} and \citet{Jin2019}. 
Column (4): Redshift from NED, measured from optical spectra. 
Column (5): Luminosity distance from \citet{Shangguan2019ApJ...870..104S}, derived from the heliocentric velocity corrected for the three-attractor flow model of \citet{Mould2000ApJ...529..786M} using $\Omega_m=0.308$, $\Omega_{\Lambda}=0.692$, and $H_0=67.8$ \kms\ Mpc$^{-1}$ \citep{Planck2016}. 
Column (6): Stellar mass from \citet{Shangguan2019ApJ...870..104S}.  
% Column (7): The total infrared luminosity in logarithm from \citet{Armus2009}.  
Column (7): \HI\ mass from the references given in Col. (10). 
Column (8): Beam size of the \HI\ observation. 
Column (9): Velocity resolution of the \HI\ line. 
Column (10): References for the \HI\ observations.
\\
%References: 
%1 \citet{Mirabel1988}, 2 \citet{Koribalski2004}, 3 \citet{Paturel2003}, 4 \cite{Courtois2009}, 5 \citet{Theureau1998}, 
%6 \citet{Springob2005}, 7 \citet{Hutchings1989}, 8 \citet{Mathewson1992}, 9 \citet{Theureau2007}, 10 \citet{Haynes2011}, 
%11 \citet{Mirabel1984}, 12 \citet{Dickel1978}, 13 \citet{Tifft1988}, 14 \citet{Roberts1978}, 15 \citet{Lewis1987}, 
%16 \citet{Sulentic1983}, 17 \citet{Peterson1974}, 18 \citet{Roth1991}, 19 \citet{Staveley-Smith1987}, 20 \citet{Bottinelli1993}.
%2 \citet{Mirabel1984}, 3 \citet{Springob2005}, 4 \citet{Hutchings1989}, 5 \citet{Dickel1978}, 
%6 \citet{Bottinelli1993}, 7 \citet{Courtois2015}, 8 \citet{Tifft1988}, 9 \citet{Roberts1978}, 10 \citet{Lewis1987}, 
%11 \citet{Sulentic1983}, 12 \citet{Theureau2007}, 13 \citet{Koribalski2004}, 14 \cite{Courtois2009}, 15 \citet{Peterson1974}, 
%16 \citet{Theureau1998}, 17 \citet{Paturel2003}, 18 \citet{Mathewson1992}, 19 \citet{Roth1991}, 20 \citet{Staveley-Smith1987}.
%*ESO 297- 11	HI confusion with PGC5959 (SO-a Prugniel1998, Sc Loveday1996) at V=5263
%*NGC 5734	HI confusion with PGC52680 (Sb RC3) at V=4141
%}
\tablerefs{
(1) \citealt{Mirabel1988ApJ...335..104M}; 
(2) \citealt{Springob2005ApJS..160..149S}; 
(3) \citealt{Theureau1998}; 
(4) \citealt{Paturel2003};  
(5) \citealt{Tifft1988ApJS...67....1T}; 
(6) \citealt{Hutchings1989AJ.....98..524H}; 
(7) \citealt{Lewis1987ApJS...63..515L}; 
(8) \citealt{Richter1991}; 
(9) \citealt{Haynes2011AJ....142..170H}; 
(10) \citealt{Dickel1978ApJ...223..391D}; 
(11) \citealt{Mirabel1984ApJ...277...92M}; 
(12) \citealt{Roberts1978AJ.....83.1026R}; 
(13) \citealt{Peterson1974AJ.....79..767P};
(14) \citealt{Mathewson1992ApJS...81..413M}; 
(15) \citealt{Koribalski2004AJ....128...16K}; 
(16) \citealt{Roth1991AJ....102.1303R}; 
(17) \citealt{Staveley-Smith1987MNRAS.224..953S}.}
% (4) \citealt{Courtois2009AJ....138.1938C}; 
% (9) \citealt{Theureau2007}; 
% (20) \citealt{Bottinelli1993}.}
\tablenotetext{a}{\HI\ spectrum strongly affected by absorption.}
}
% (1) \citealt{Lelli2016AJ....152..157L};
% (2) \citealt{Swaters2009AA...493..871S}.}
\label{tbl:achv-list}
\end{deluxetable*}

\begin{figure*}[hbt!]
\setcounter{figure}{2}
\epsscale{0.45}
\plotone{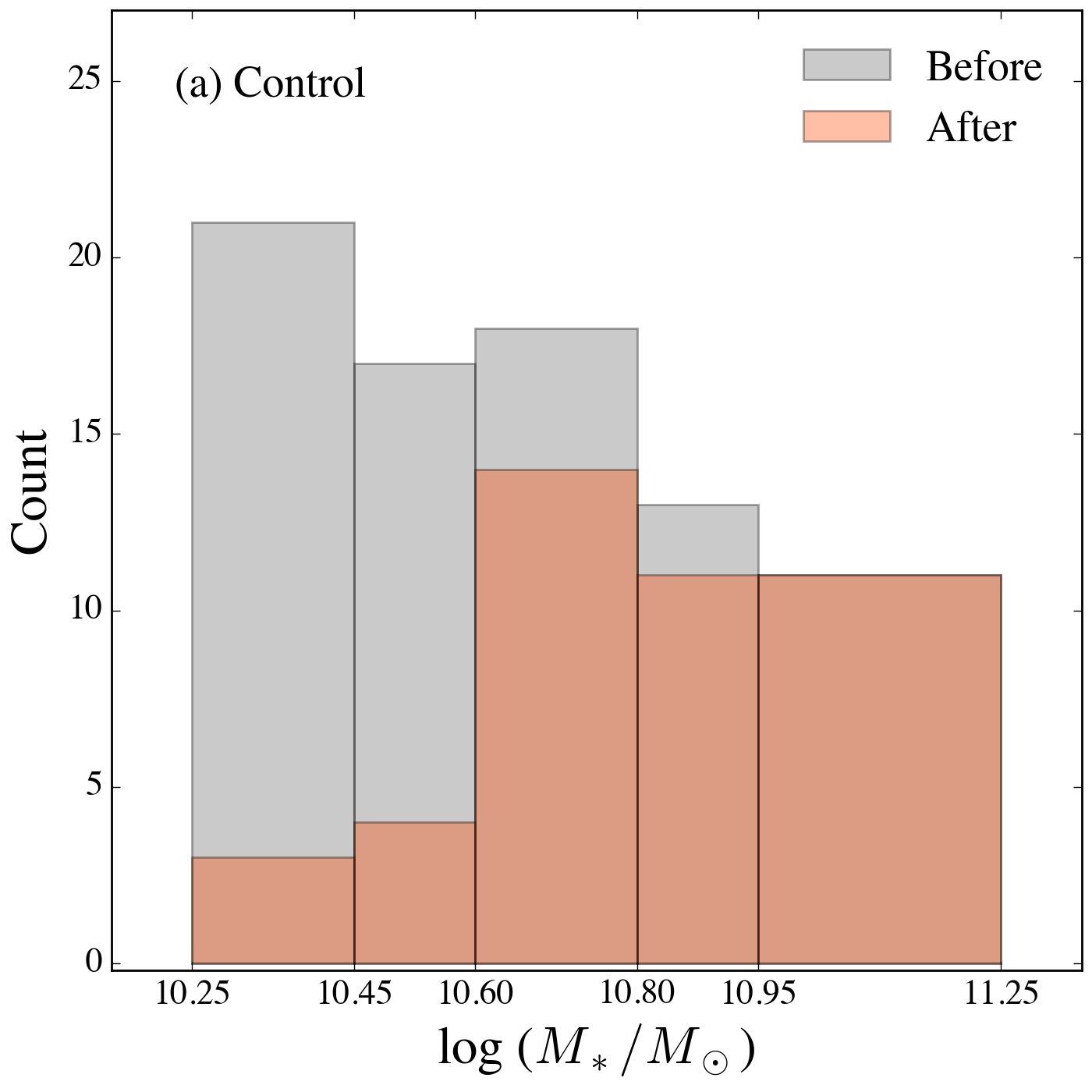}
\epsscale{0.45}
\plotone{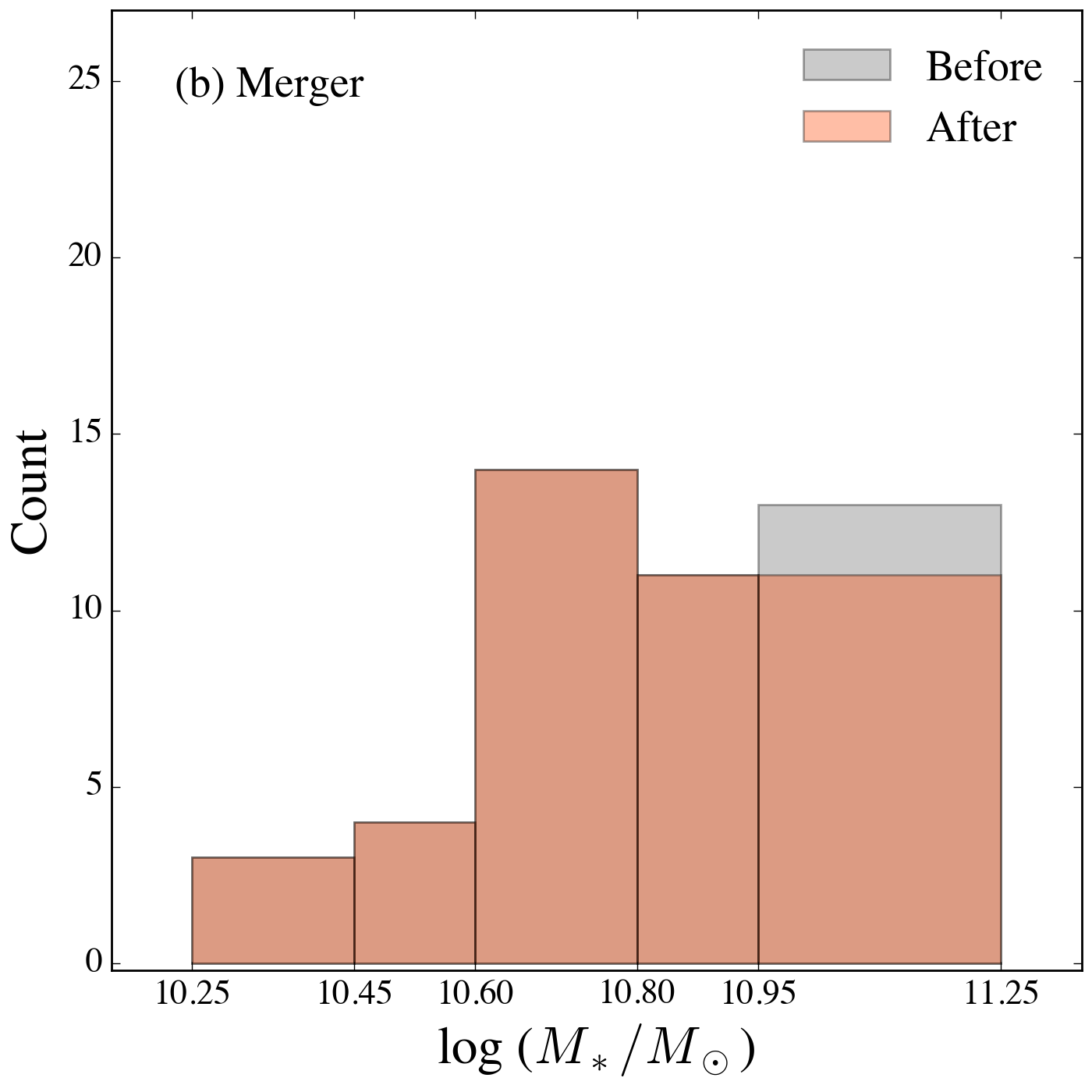}
% \epsscale{0.38}
% \plotone{Hist_lgMSMtchHtch.png}
\caption{Stellar mass distribution of the (a) control and (b) merger galaxies before (gray) and after (orange) the one-to-one sample matching.  Note that the mass bins are uneven.  
}
\label{fig:Hist_MStar}
\end{figure*}
%edited by LCH 2021.06.17
%edited by LCH 2021.11.29

% \FloatBarrier
\begin{figure*}[hbt!]
\setcounter{figure}{3}

\epsscale{0.45}
\plotone{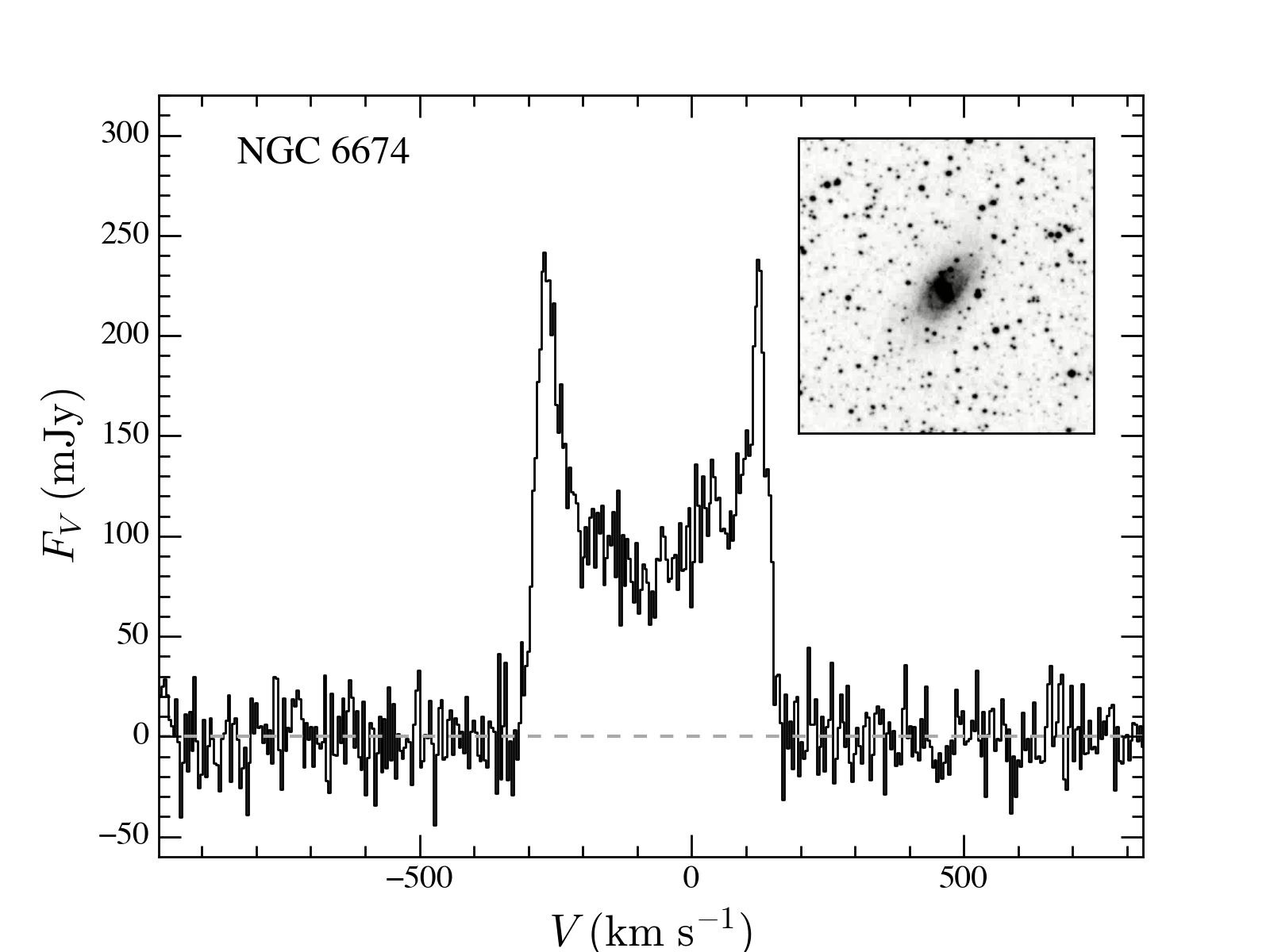}
\epsscale{0.45}
\plotone{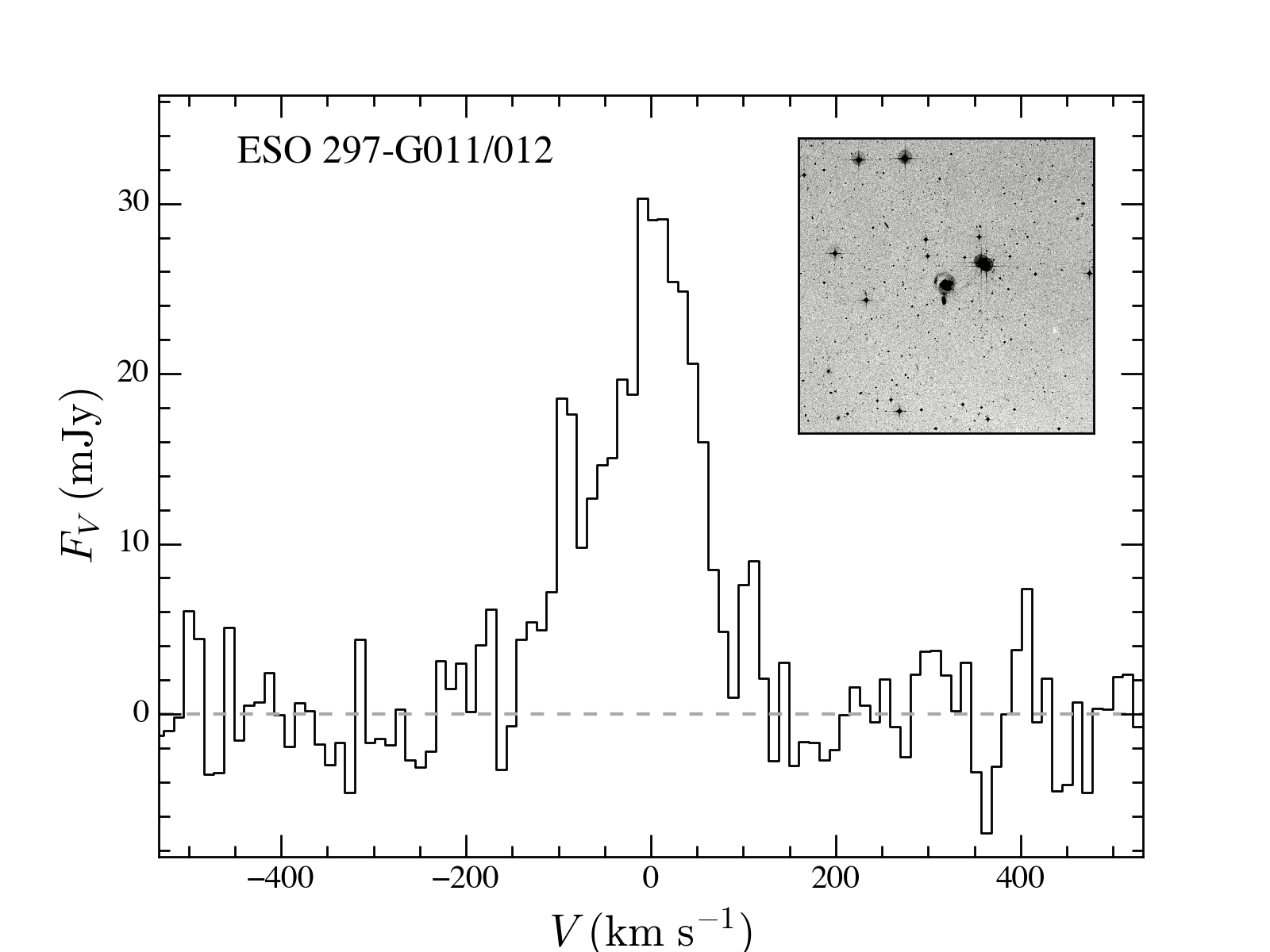}

\epsscale{0.45} %**
\plotone{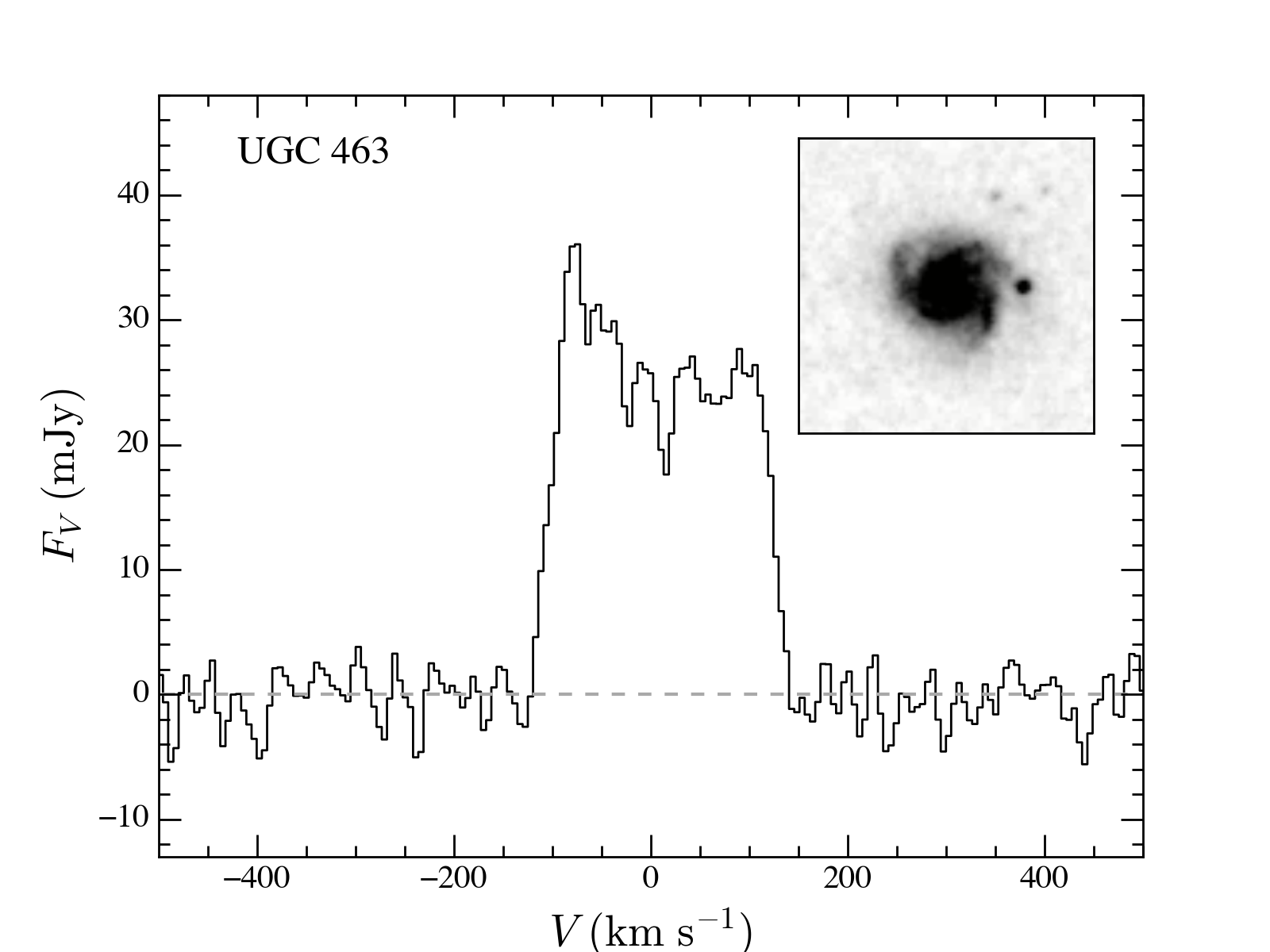}
\epsscale{0.45}
\plotone{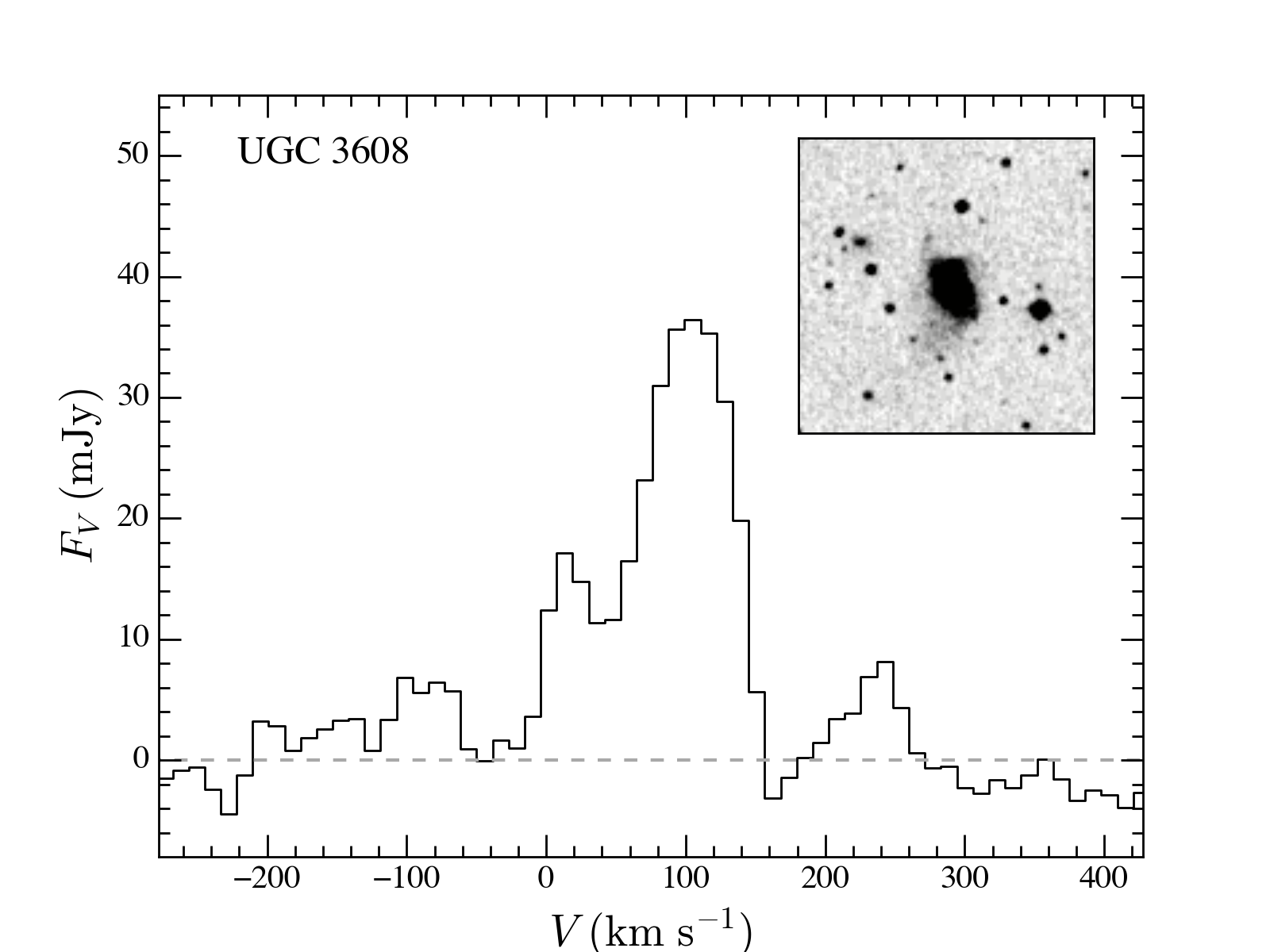}

\epsscale{0.45}
\plotone{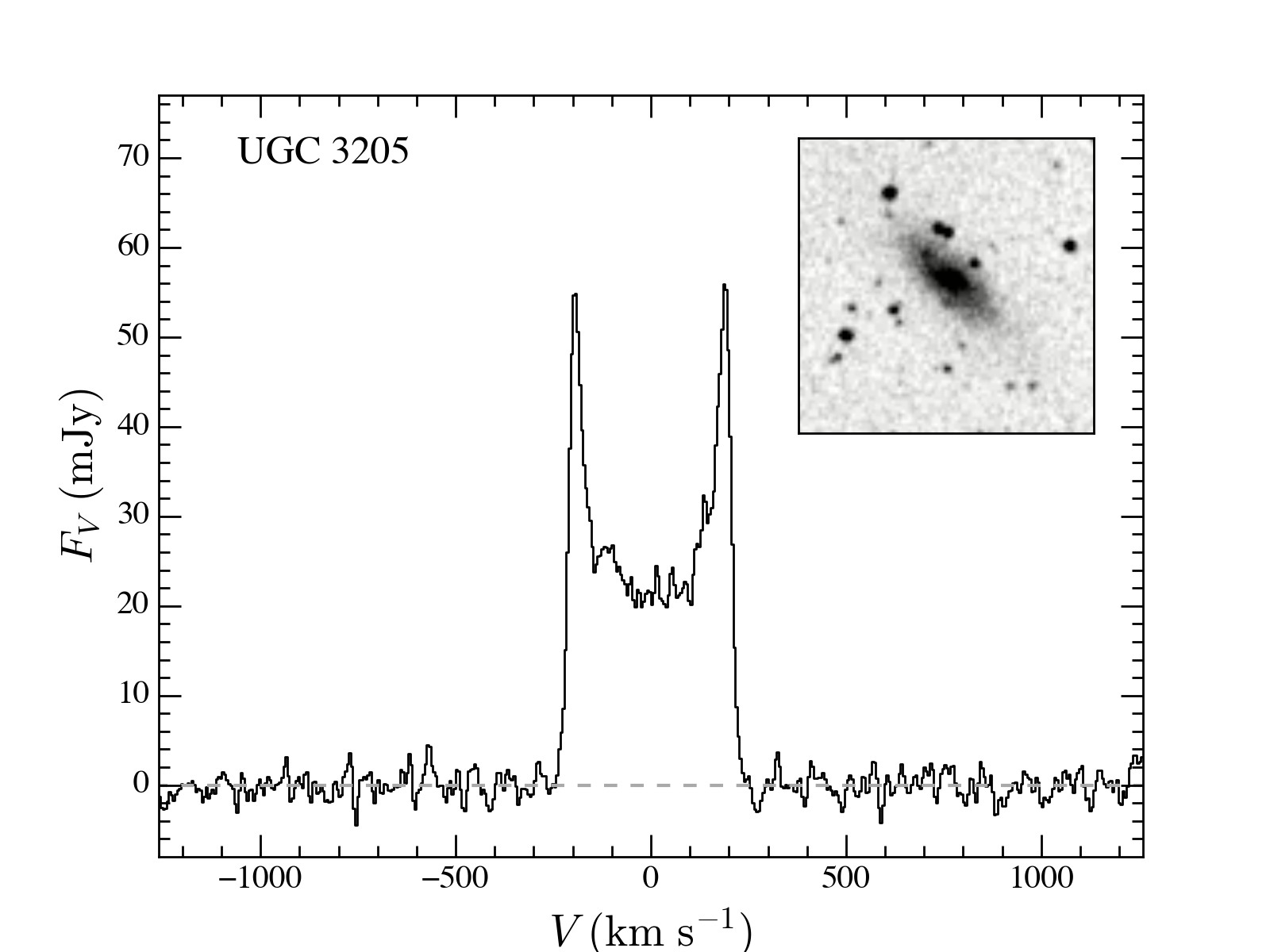}
\epsscale{0.45}
\plotone{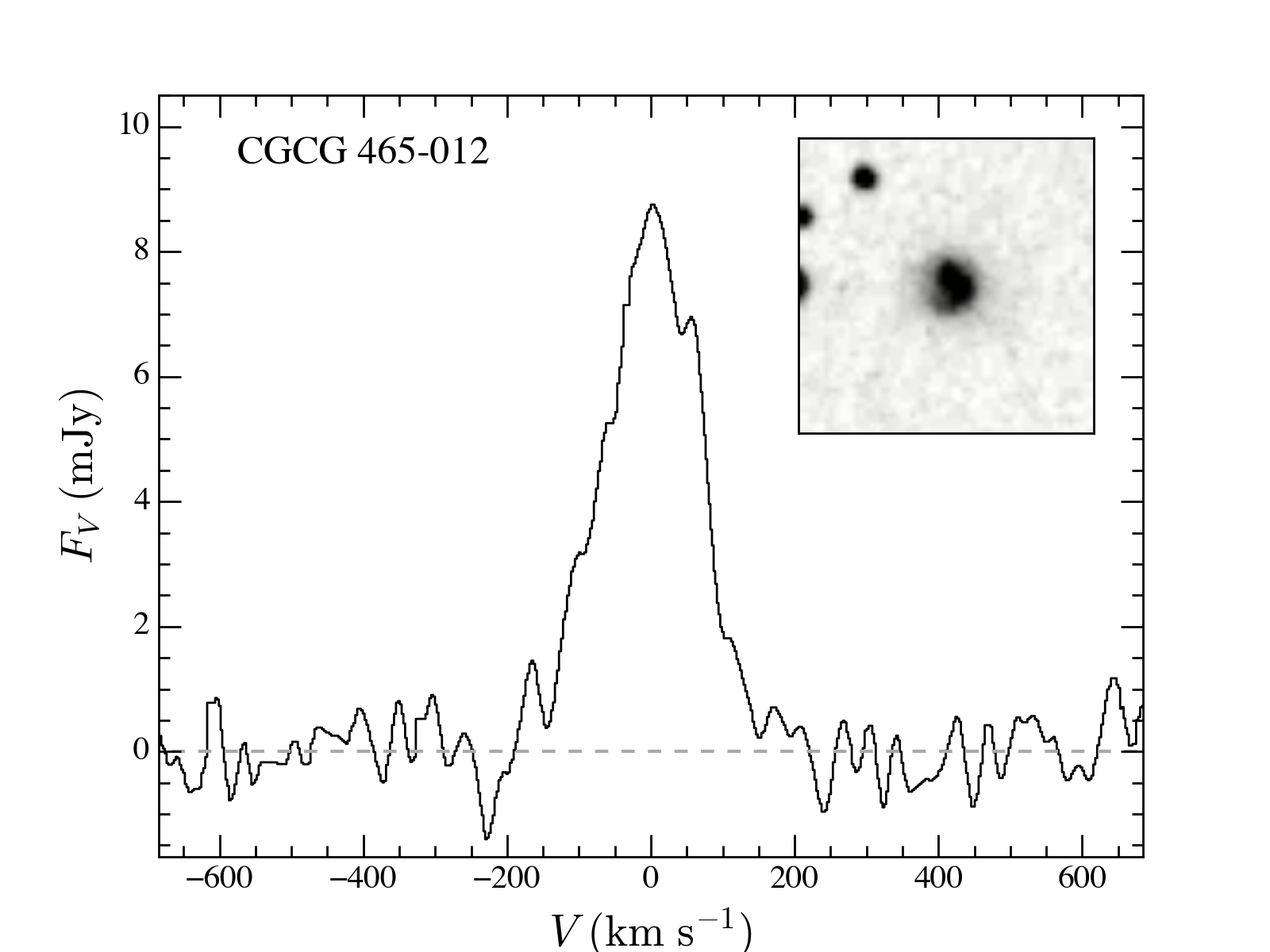}

\epsscale{0.45}
\plotone{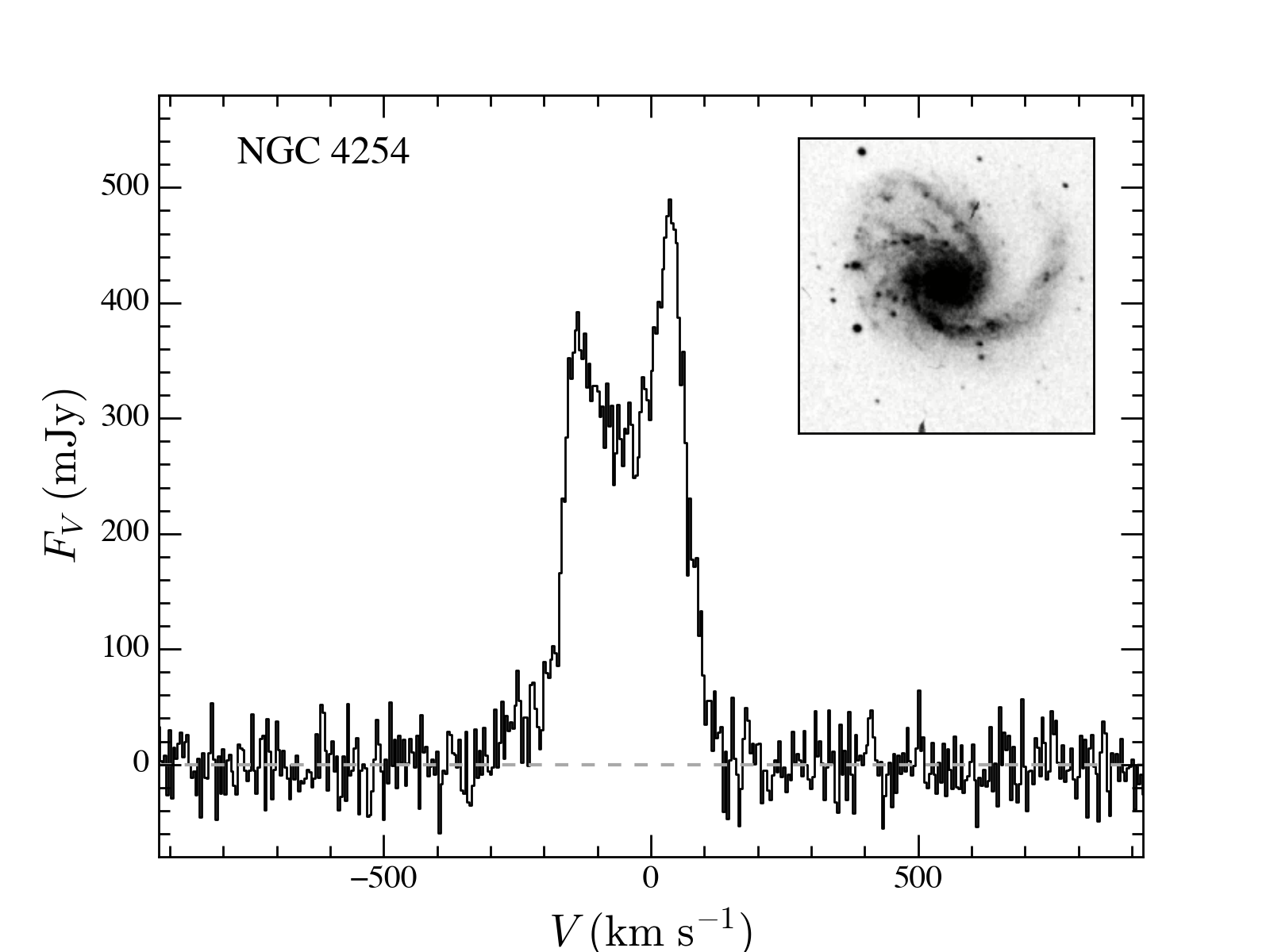}
\epsscale{0.45}
\plotone{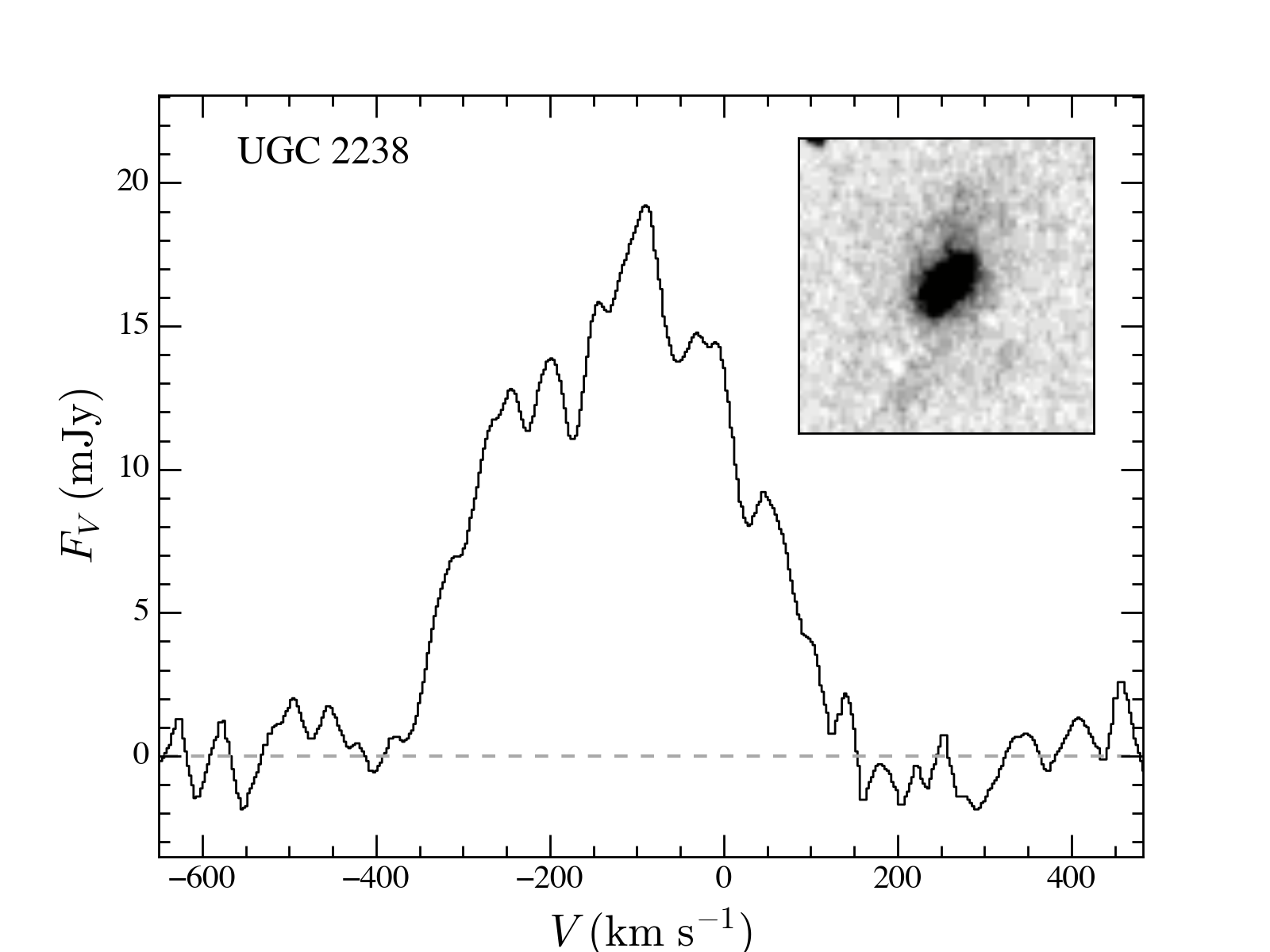}

\caption{\HI\ profiles of example (left) non-merger and (right) merger galaxies.  The inset in each panel shows the optical (4680~\r{A} for ESO~297$-$G011, 6450~\r{A} for the other galaxies) Digitized Sky Survey image from NED.  The two galaxies in each row are matched in stellar mass.  The spectra for the full sample are given in the figure set in the electronic version of the paper.  
}
\label{fig:opt-hi}
\end{figure*}

\begin{figure*}[ht!]
\epsscale{0.65}
\plotone{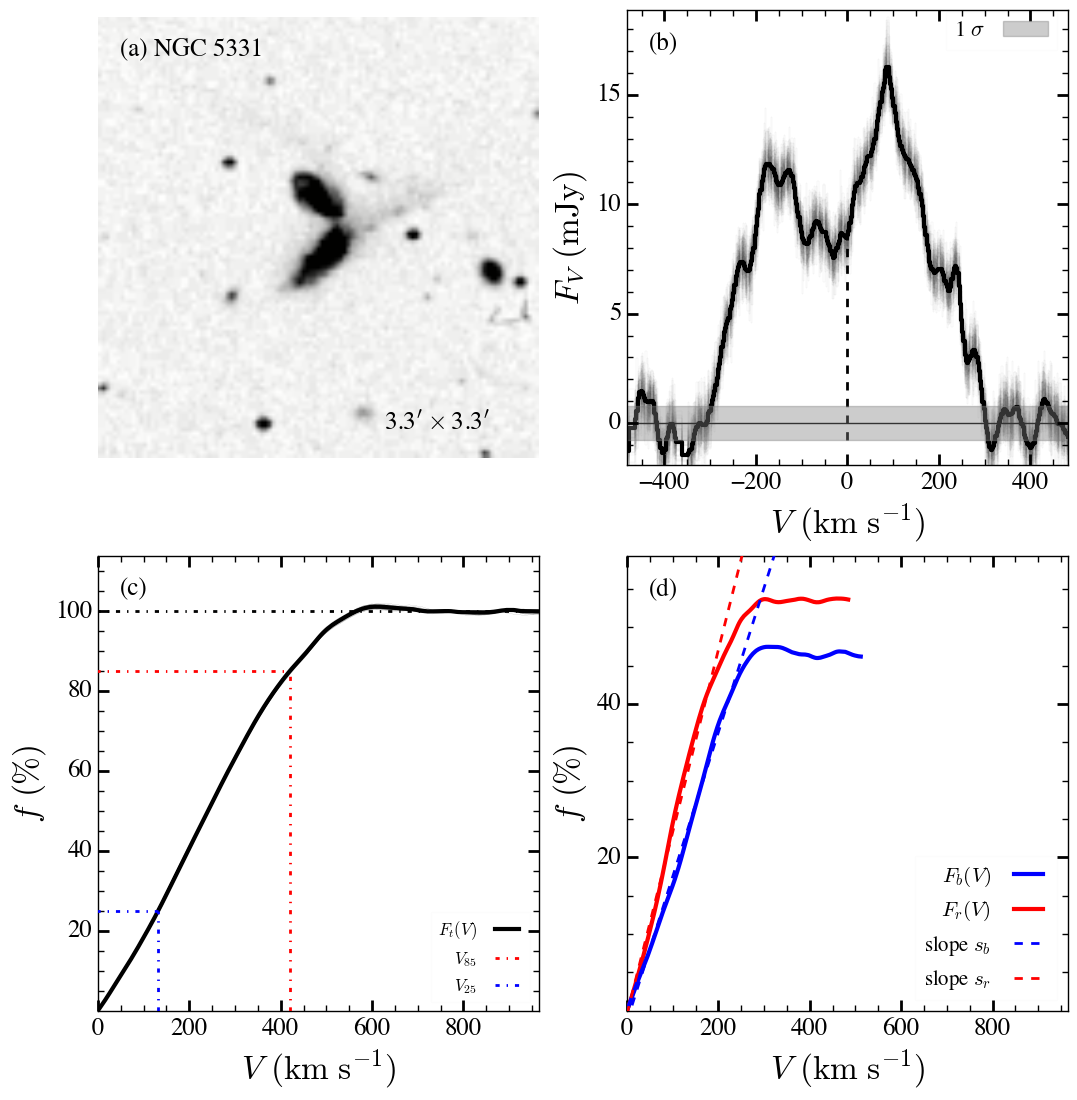}
\caption
{Illustration of our analysis technique for the merger galaxy NGC~5331.  Panel (a) shows the optical (4680~\r{A}) Digitized Sky Survey image from NED, whose size is given on the lower-right corner.  Panel (b) displays the \HI\ spectrum (thick black line) on a velocity scale centered in the rest frame; the grey shaded region shows the $1\,\sigma$ uncertainty of the flux intensity.  Panel (c) presents the normalized CoG for the total spectrum $f = F_t(V)/F$ (black solid line), which converges at $f=100\%$ (black dot-dashed line).  The colored dot-dashed lines mark the line widths at 25\% (blue: $V_{25}$) and 85\% (red: $V_{85}$) of the total integrated flux.  Panel (d) shows the normalized CoG for the blue side of the spectrum $f=F_b(V)/F$ (solid blue) and the red side of the spectrum $f=F_r(V)/F$ (solid red). The linear fits of the rising part of the CoG are shown as blue (slope $s_b$) and red (slope $s_r$) dashed lines. The thin black lines in panels (b) and (c) are the results from 50 sets of Monte Carlo simulations. }
\label{fig:cog}
\end{figure*}

\begin{figure*}[ht!]
\epsscale{0.65}
\plotone{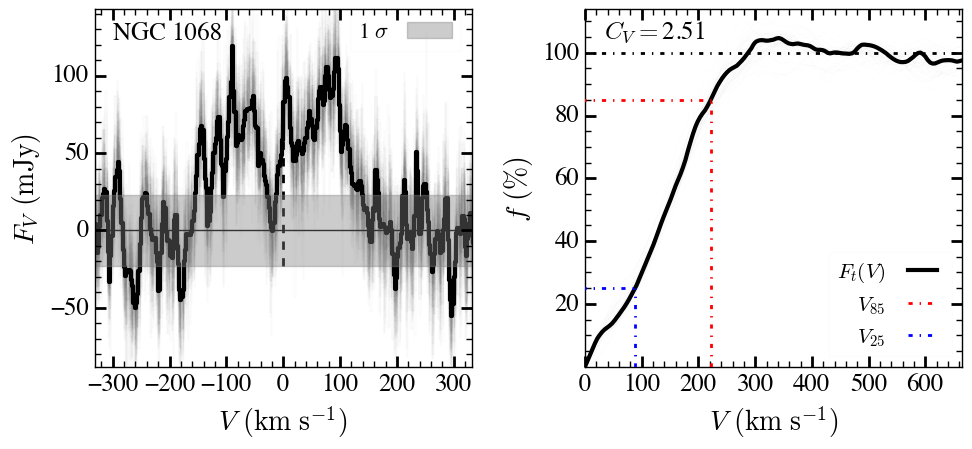}
\epsscale{0.65}
\plotone{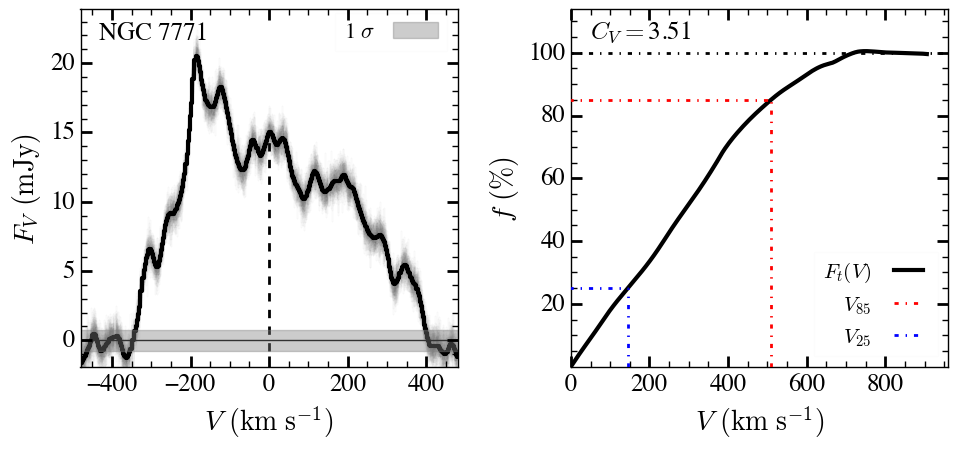}
\epsscale{0.65}
\plotone{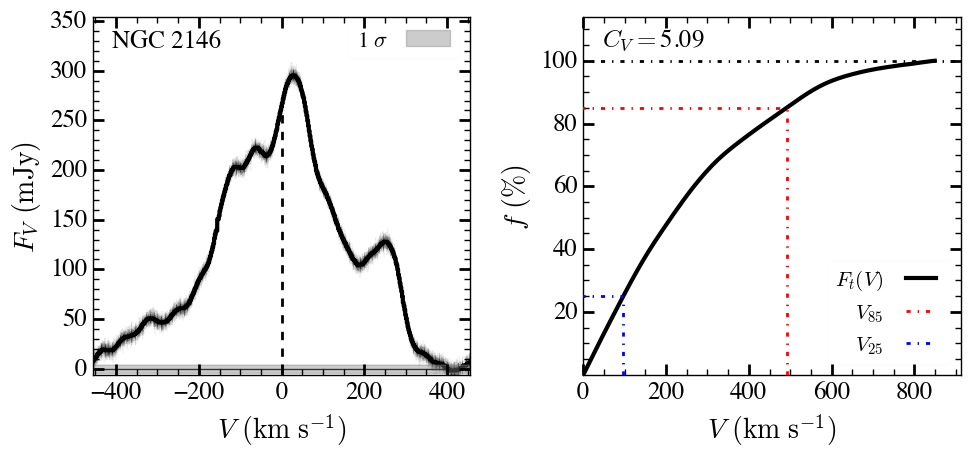}
\caption{
The \HI\ spectra (left) and their CoGs (right) of three galaxies with different degrees of profile concentration, $C_V \equiv V_{85}/V_{25}$.  The curves and lines are the same as in Figure~\ref{fig:cog}. 
}
\label{fig:cv-trend}
\end{figure*}

\begin{figure*}[ht!]
\epsscale{0.65}
\plotone{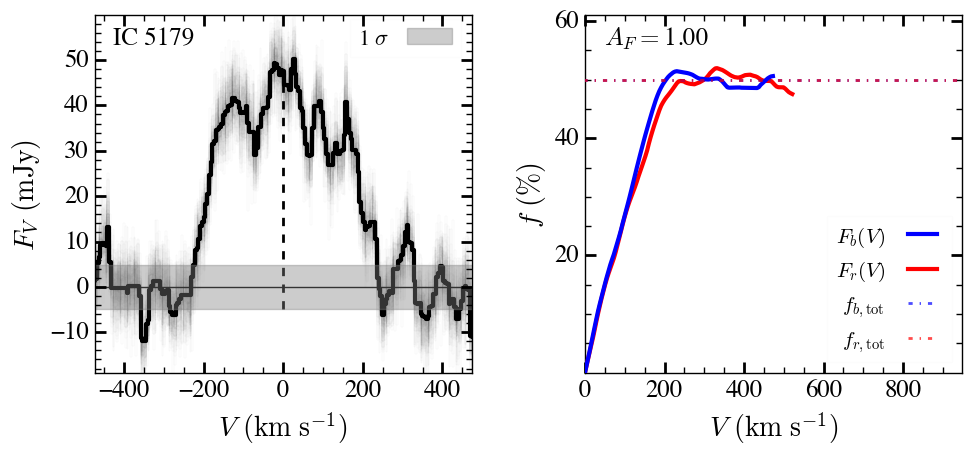}
\epsscale{0.65}
\plotone{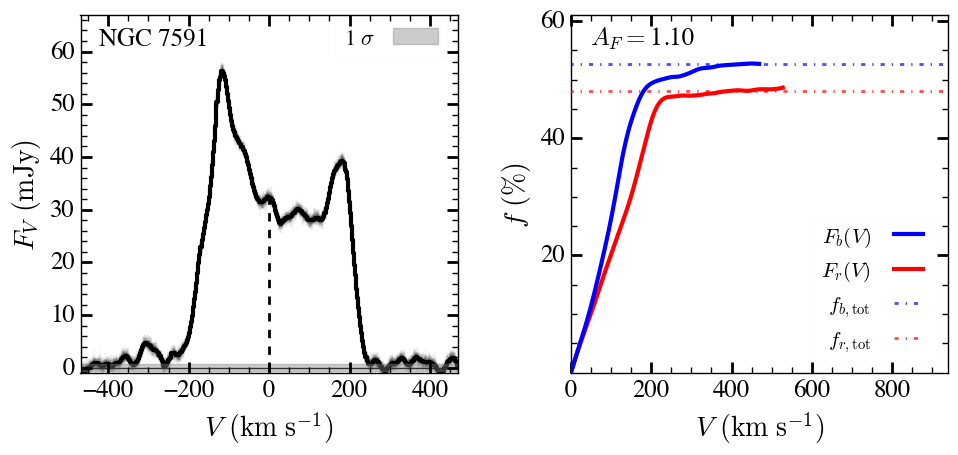}
\epsscale{0.65}
\plotone{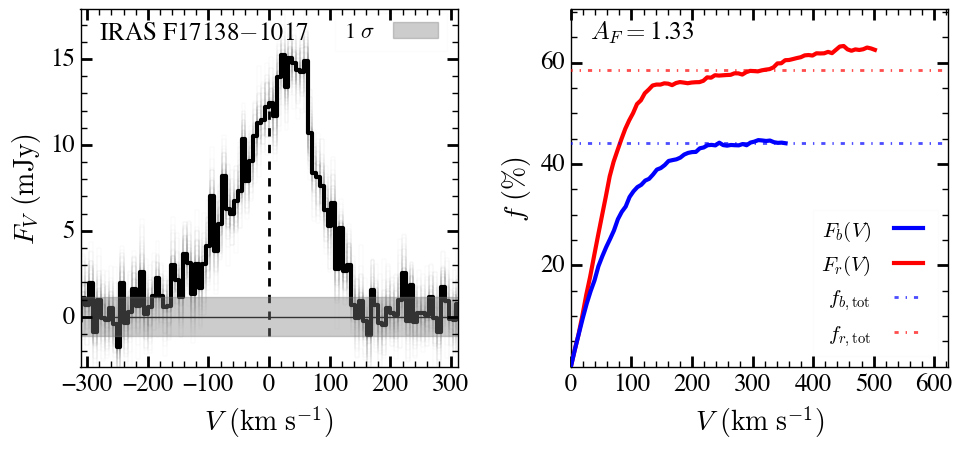}
\caption{
The \HI\ spectra (left) and their CoGs (right) for the blueshifted (blue) and redshifted (red) sides of three galaxies with different degrees of flux asymmetry, $A_F \equiv f_{b, \rm tot}/f_{r, \rm tot}$, where the cumulative flux fraction of the blue ($f_{b, \rm tot}$) and red ($f_{r, \rm tot}$) sides of the profile are given by the horizontal blue and red dot-dashed lines, respectively.
}
\label{fig:af-trend}
\end{figure*}

\begin{figure*}[ht!]
\epsscale{0.65}
\plotone{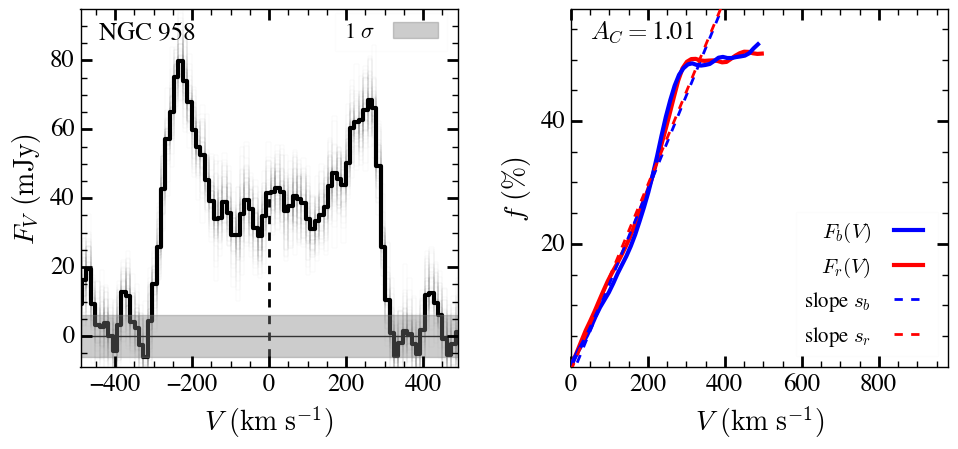}
\epsscale{0.65}
\plotone{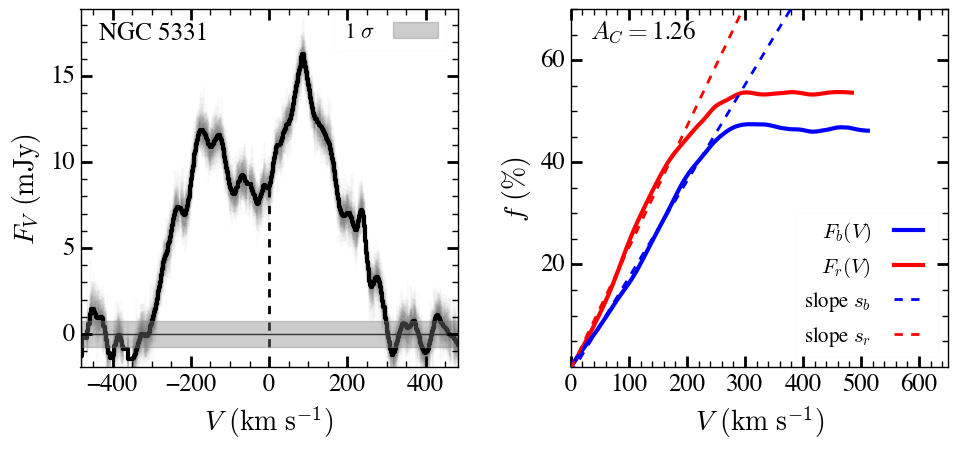}  
\epsscale{0.65}
\plotone{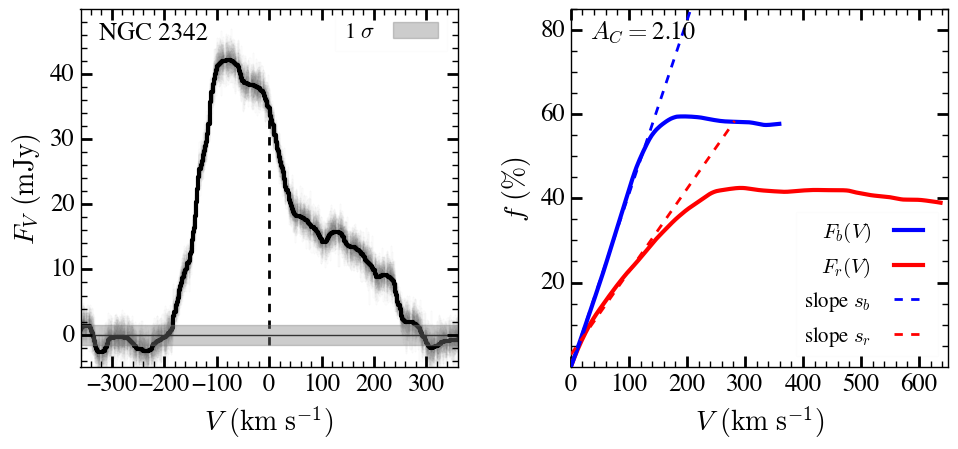}
\caption{
The \HI\ spectra (left) and their CoGs (right) for the blueshifted (blue) and redshifted (red) sides of three galaxies with different degrees of flux distribution asymmetry, $A_C \equiv s_{b}/s_{r}$, where the slope of the rising part of the CoG of the blue ($s_{b}$) and red ($s_{r}$) sides of the profile are given by the blue and red dashed lines, respectively.
}
\label{fig:ac-trend}
\end{figure*}

\section{\HI\ Profile Measurements} 
\label{sec:CoG}

We quantify the \HI\ profiles of the merger and control samples using the CoG method of \cite{Yu2020ApJ...898..102Y}.  Here, we briefly summarize the main points, which are also illustrated for an example galaxy in Figure~\ref{fig:cog}.  After subtracting the baseline and masking channels with obvious contamination by RFI, we search for \HI\ emission in channels within a range $\pm 500$~\kms around the optical systemic velocity of the galaxy.  Spectral segments with at least three consecutive channels are selected if their mean flux density is above 0.7 times the maximum flux density of the spectrum.  The lowest and highest velocity channels of the selected segments determine the velocity range of the line emission.  The flux intensity-weighted velocity of the line-emitting channels defines the central velocity $V_c$ (Figure~\ref{fig:cog}b).  Starting at velocity $V_c$, we integrate the \HI\ flux toward both sides of the profile to construct the CoG of the line (Figure~\ref{fig:cog}c).  If the baseline of the spectrum has been measured properly and subtracted, the CoG rises monotonically until it reaches a plateau.  The total line flux $F$ is the median integrated flux on the flat part of the CoG.  The line widths $V_{25}$ and $V_{85}$ are the velocity widths enclosing 25\% and 85\% of the total flux (blue and red dash-dotted lines).  Defining the concentration of the line as $C_V \equiv V_{85}/V_{25}$, the line profile varies from double-horned to single-peaked as $C_V$ increases (Figure~\ref{fig:cv-trend}).  To quantify the asymmetry of the line, we calculate the CoG for the blue and red sides of the profile (red and blue curves in Figure~\ref{fig:cog}d).  The asymmetry parameter $A_F$ is defined as the larger of the ratios of the integrated flux of one side of the line relative to the other (Figure~\ref{fig:af-trend}). The growth rate of the CoG of the two sides of the line also reflects the line asymmetry: $A_C$ denotes the larger of the ratios of the slopes of the rising part of the CoG of the two sides (Figure~\ref{fig:ac-trend}). The parameters $A_F$ and $A_C$ are always equal to or larger than 1.  These newly defined profile asymmetry and concentration parameters are stable against conditions of relatively low signal-to-noise ratio (${\rm S/N} \lesssim 10$; \citealt{Yu2020ApJ...898..102Y}).

Following \citet{Yu2020ApJ...898..102Y}, we obtain intrinsic line widths from the observed line widths after correcting them for instrumental resolution, redshift, and turbulent motions.  The profile asymmetry and concentration parameters are corrected for the effect of S/N.  The uncertainties of the measurements are the quadrature sum of statistical uncertainties estimated from Monte Carlo realizations that add Gaussian noise to the observed spectrum and systematic uncertainties from mock simulations \citep{Yu2020ApJ...898..102Y}.  The final measured parameters are given in Table~\ref{tbl:paras-list} for the merger sample and in Table \ref{tbl:paras-list_1} for the control sample.

%edited by LCH 2021.06.08
%edited by LCH 2021.11.29

% \startlongtable
% \begin{deluxetable}{cD@{$\pm$}DD@{$\pm$}DD@{$\pm$}DD@{$\pm$}DD@{$\pm$}DD@{$\pm$}DDDc}
\begin{deluxetable*}{cc@{ $\pm$ }cc@{ $\pm$ }cc@{ $\pm$ }cc@{ $\pm$ }cc@{ $\pm$ }cc@{ $\pm$ }cccc}[ht!]
\tablenum{3}
\centering
\small\addtolength{\tabcolsep}{-0.5pt}
%\tabletypesize{\small}
\tabletypesize{\small}
\tablecolumns{15}
\tablewidth{0pt} 
\tablecaption{Physical Parameters of Merger Sample Derived from the \HI\ Spectra}
\tablehead{
\colhead{Galaxy} &
\multicolumn2c{$V_c$} &
\multicolumn2c{$F$} &
\multicolumn2c{$V_{85}$} &
\multicolumn2c{$A_F$} &
\multicolumn2c{$A_C$} &
\multicolumn2c{$C_V$} &
\colhead{S/N} &
\colhead{log \MHI} &
\colhead{Notes} \\ 
\colhead{} &     
\multicolumn2c{(\kms)} & 
\multicolumn2c{(Jy\ \kms)} &
\multicolumn2c{(\kms)} &
\multicolumn2c{} &
\multicolumn2c{} & 
\multicolumn2c{} &  
\colhead{} & 
\colhead{(\msun)} &  
\colhead{}
\\
\colhead{(1)} &
\multicolumn2c{(2)} &
\multicolumn2c{(3)} &
\multicolumn2c{(4)} &
\multicolumn2c{(5)} &
\multicolumn2c{(6)} &
\multicolumn2c{(7)} &
\multicolumn1c{(8)} &
\multicolumn1c{(9)} &
%\multicolumn4c{(13)}&
\colhead{(10)}  
}
\decimals % to align according to decimals (column identifier: D)
\startdata  % 75 objects
CGCG~043-099 & 11237 & 3 & 2.68 & 0.02 & 263 & 16 & 1.13 & 0.08 & 1.12 & 0.11 & 3.88 & 0.05 & 43.7 & 10.31 & \nodata \\
CGCG~052-037 &  7398 & 3 & 1.15 & 0.01 & 253 & 15 & 1.06 & 0.08 & 1.09 & 0.11 & 4.41 & 0.07 & 37.4 & 9.59 & 2 \\
CGCG~453-062 & 7453 & 3 & 1.22 & 0.01 & 338 & 21 & 1.03 & 0.08 & 1.25 & 0.13 & 3.67 & 0.07 & 23.7 & 9.56 & 2 \\
CGCG~465-012 & 6659 & 3 & 1.37 & 0.01 & 180 & 11 & 1.07 & 0.08 & 1.06 & 0.11 & 4.51 & 0.06 & 32.8 & 9.49 & \nodata \\
ESO~173-G015 & 2926 & 13 & 28.60 & 0.83 & 242 & 83 & 1.00 & 0.27 & 1.08 & 1.00 & 3.15 & 0.12 & 8.9 & 9.92 & 1 \\
ESO~297-G011/012 & 5187 & 13 & 3.90 & 0.12 & 167 & 57 & 1.06 & 0.29 & 1.39 & 1.29 & 4.24 & 0.26 & 8.6 & 9.73 & \nodata \\
\enddata
\tablecomments{
Column (1): Galaxy name. 
Column (2): Flux intensity-weighted central velocity. 
Column (3): Total integrated flux of \HI\ line.
Column (4): Corrected line width measured at 85\% of the total flux. 
Columns (5) and (6): Flux asymmetry and flux distribution asymmetry, corrected for S/N as described in Section~\ref{sec:CoG}.
Column (7): Corrected concentration of the \HI\ profile. 
Column (8): S/N of the profile, calculated based on Equation (5) in \citet{Yu2020ApJ...898..102Y}.
Column (9): \HISS\ mass; for an assumed uncertainty of 10\% for the distance and 15\% for the \HI\ flux, the typical uncertainty of $\log$\MHI\ is 0.11 dex. 
Column (10): Notes: 1 = mask generated for the spectrum; 2 = baseline subtracted in this work.  
}
\label{tbl:paras-list}
\end{deluxetable*}

%edited by LCH 2021.11.29

{\color{red}
\begin{deluxetable*}{cc@{ $\pm$ }cc@{ $\pm$ }cc@{ $\pm$ }cc@{ $\pm$ }cc@{ $\pm$ }cc@{ $\pm$ }cccc}[ht!]
\tablenum{4}
\centering
\small\addtolength{\tabcolsep}{-0.5pt}
%\tabletypesize{\small}
\tabletypesize{\small}
\tablecolumns{15}
\tablewidth{0pt} 
\tablecaption{Physical Parameters of the Control Sample Derived from the \HI\ Spectra}
\tablehead{
\colhead{Galaxy} &
\multicolumn2c{$V_c$} &
\multicolumn2c{$F$} &
\multicolumn2c{$V_{85}$} &
\multicolumn2c{$A_F$} &
\multicolumn2c{$A_C$} &
\multicolumn2c{$C_V$} &
\colhead{S/N} &
\colhead{log \MHI} &
\colhead{Notes} \\ 
\colhead{} &     
\multicolumn2c{(\kms)} & 
\multicolumn2c{(Jy\ \kms)} &
\multicolumn2c{(\kms)} &
\multicolumn2c{} &
\multicolumn2c{} & 
\multicolumn2c{} &  
\colhead{} & 
\colhead{(\msun)} &  
\colhead{}
\\
\colhead{(1)} &
\multicolumn2c{(2)} &
\multicolumn2c{(3)} &
\multicolumn2c{(4)} &
\multicolumn2c{(5)} &
\multicolumn2c{(6)} &
\multicolumn2c{(7)} &
\multicolumn1c{(8)} &
\multicolumn1c{(9)} &
%\multicolumn4c{(13)}&
\colhead{(10)}  
}
\decimals % to align according to decimals (column identifier: D)
\startdata  % 75 objects
AGC~190187 &  8060 &   1 &    1.17 &   0.23 &  271 &    3 &  1.02 &   0.27 &  1.04 &   0.92 &  3.09 &   0.08 &    9.3 &   9.56 &           2 \\
AGC~260268 &  8950 &   1 &    1.75 &   0.28 &  165 &    3 &  1.05 &   0.07 &  1.23 &   1.08 &  2.72 &   0.07 &   13.7 &   9.88 &                     \nodata \\
IC~381 &  2485 &   1 &   22.11 &   3.50 &  240 &    1 &  1.12 &   0.08 &  1.33 &   1.23 &  2.66 &   0.01 &  166.4 &   9.91 &                 2 \\
IC~529 &  2272 &   1 &   27.93 &   4.42 &  272 &    1 &  1.03 &   0.07 &  1.13 &   1.04 &  2.91 &   0.01 &  148.5 &   9.90 &                  1 \\
IC~1074 &  7791 &  18 &    1.21 &   0.24 &  470 &   30 &  1.16 &   0.28 &  1.08 &   0.94 &  2.55 &   0.41 &    5.5 &   9.61 &                     \nodata \\
IC~2387 &  7686 &   1 &    4.94 &   0.78 &  311 &    2 &  1.11 &   0.08 &  1.04 &   0.93 &  2.92 &   0.05 &   19.1 &  10.26 &                     \nodata \\
UGC~448 &  4859 &   1 &    6.23 &   0.99 &  157 &    1 &  1.01 &   0.07 &  1.05 &   0.96 &  3.09 &   0.05 &   27.6 &   9.74 &                     \nodata \\
NGC~514 &  2476 &   1 &   28.10 &   4.45 &  228 &    1 &  1.06 &   0.07 &  1.02 &   0.93 &  2.63 &   0.03 &   28.5 &   9.86 &                  1 \\
UGC~463 &  4451 &   1 &    6.02 &   0.95 &  192 &    2 &  1.10 &   0.08 &  1.16 &   1.06 &  3.05 &   0.06 &   23.1 &   9.42 &                     \nodata \\
UGC624 &  4770 &   1 &   11.68 &   1.85 &  467 &    1 &  1.04 &   0.07 &  1.04 &   0.96 &  2.87 &   0.02 &   43.7 &  10.16 &                     \nodata \\
\enddata
\tablecomments{
Column (1): Galaxy name. 
Column (2): Flux intensity-weighted central velocity. 
Column (3): Total integrated flux of \HI\ line.
Column (4): Corrected line width measured at 85\% of the total flux. 
Columns (5) and (6): Flux asymmetry and flux distribution asymmetry, corrected for S/N as described in Section~\ref{sec:CoG}.
Column (7): Corrected concentration of the \HI\ profile. 
Column (8): S/N of the profile, calculated based on Equation (5) in \citet{Yu2020ApJ...898..102Y}.
Column (9): \HISS\ mass; for an assumed uncertainty of 10\% for the distance and 15\% for the \HI\ flux, the typical uncertainty of $\log$\MHI\ is 0.11 dex. 
Column (10): Notes: 1 = mask generated for the spectrum; 2 = baseline subtracted in this work.  
}
\label{tbl:paras-list_1}
\end{deluxetable*}
}

\section{Comparison Between Merger and Non-merger Galaxies}
\label{sec:4}

\begin{figure*}
\plotone{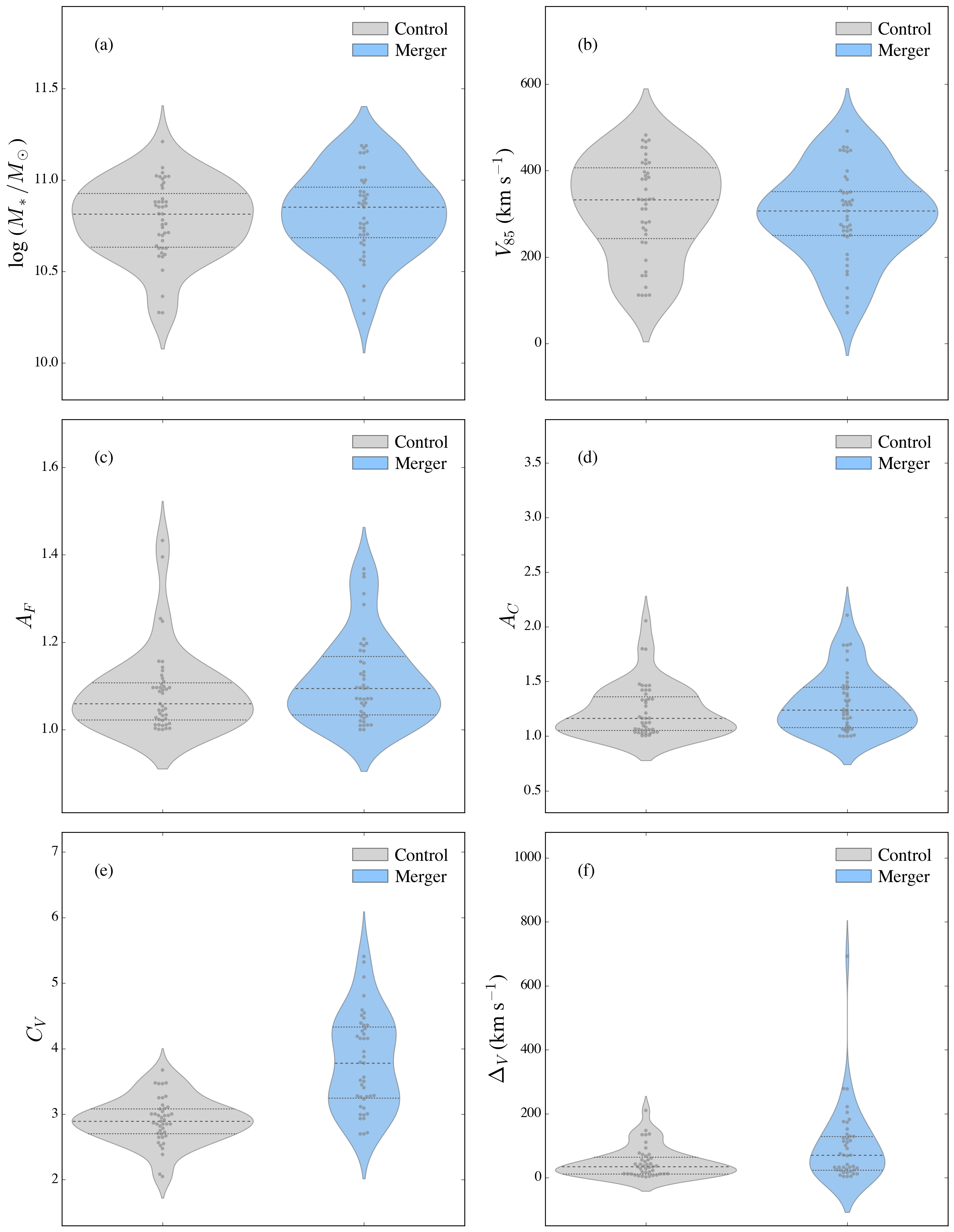}
\caption{
Violin plots of the 50th percentile distributions of (a) $\log M_*$, (b) $V_\mathrm{85}$, (c) $A_F$, (d) $A_C$, (e) $C_V$, and (f) $\Delta V$, for non-merger, control (grey) and merger (blue) galaxies.  The distributions were derived from 1000 realizations of K-S tests.  The grey dashed lines show the median of the points, while the dotted grey lines are the 25th and 75th percentile of the distribution.  The area of each violin is the same, and the width is scaled by the number of points in that bin.  The dark grey ``beeswarm'' points on the violins represent the distribution of values.  The contour is a kernel density estimation showing the distribution shape of the data. Wider sections represents a higher probability, while the narrower sections represent a lower probability.
}
\label{fig:vln-dis}
\end{figure*}

\begin{figure}
\includegraphics[height=0.30\textheight]{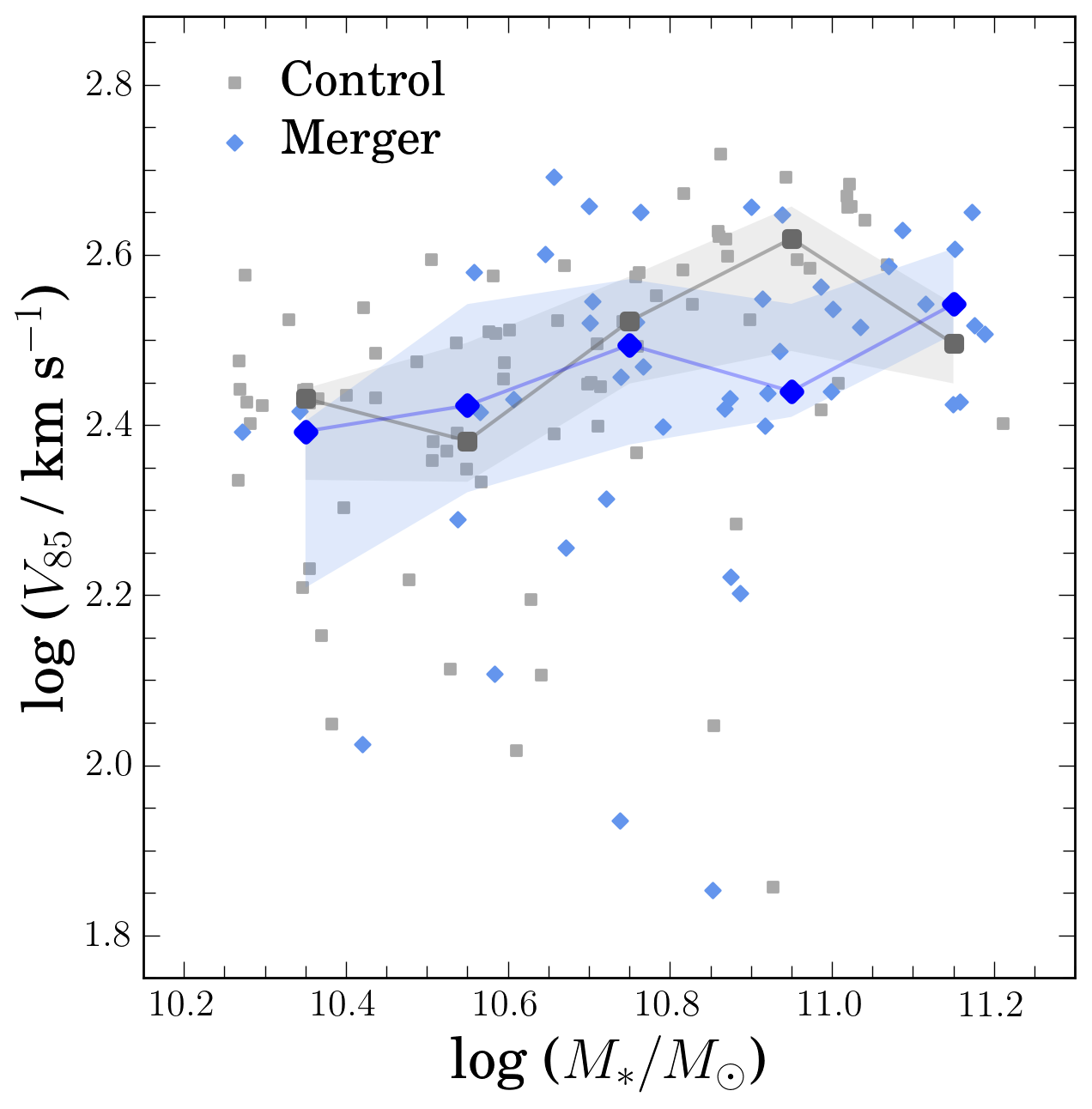}
\caption{The variation of rotation velocity ($V_{85}$) with stellar mass ($M_*$) of the merger (blue diamond) and control (grey square) samples. The large symbols and the corresponding connecting lines give the median values of the objects in bins of 0.2 dex in $\log (M_*/M_\odot)$ from 10.25 to 11.25, and the shaded regions are the 25th and 75th percentiles of the distribution.}
\label{fig:V85_M}
\end{figure}

With the \HI\ profile measurements in hand, we compare systematically the statistical properties of the merger and non-merger galaxies.  The violin plots of Figure~\ref{fig:vln-dis} display the distributions of the main parameters for the two samples: $M_*$, $V_\mathrm{85}$, $A_F$, $A_C$, $C_V$, and $\Delta V$. The violin plot is similar to a box plot, a non-parametric, visual representation of a variable distribution across different samples, but includes the kernel density estimate of the sample distribution (see the caption of Figure~\ref{fig:vln-dis} for more details). Matched in stellar mass by design (Figure~\ref{fig:vln-dis}a), the two samples exhibit statistically similar rotation velocities $V_{85}$ (Figure~\ref{fig:V85_M})\footnote{Because the inclination angle is difficult to measure or poorly defined in mergers, in this paper the quantity $V_{85}$ pertains to the projected rotation velocity, for both the merger and non-merger samples.}.  In other words, mergers cannot be distinguished from non-mergers in terms of their \HI\ line widths. This is somewhat surprising, as one naively might have thought that merging systems, in the aftermath of two galaxies colliding, would show a larger range of integrated velocities. This expectation holds only for one out of the five mass bins ($M_* = 10^{10.85}-10^{11.05}\,M_\odot$) in Figure~\ref{fig:V85_M}, for which mergers have a median $V_{85} = 270\,{\rm km\,s^{-1}}$, to be compared to $V_{85} = 397\,{\rm km\,s^{-1}}$ for the control sample.  The violin density plots of the two groups are nearly identical (Figure~\ref{fig:vln-dis}b).  Based on the Kolmogorov-Smirnov (K-S) test with a $p$-value = $0.36^{+0.02}_{-0.01}$ (Table~\ref{tbl:ks-rst}), the null hypothesis that the two samples are drawn from the same parent distribution cannot be rejected. \footnote{Following convention, we consider two distributions significantly different if $p<0.05$.}

If mergers do not produce notably broader lines, what imprint do they leave on their line asymmetry?  This comes as an even bigger surprise.  Neither their distribution of $A_F$ nor $A_C$ stands out compared to that of non-mergers (Figures~\ref{fig:vln-dis}c and \ref{fig:vln-dis}d).  According to the K-S test, the asymmetry parameters of merger and control galaxies are drawn from a similar distribution, with $p = 0.40^{+0.02}_{-0.02}$ for $A_F$ and $p = 0.48^{+0.01}_{-0.02}$ for $A_C$. Taken at face value, our results are at odds with those of \citet{Bok2019MNRAS.484..582B}, whose analysis of \HI\ asymmetry of $\sim 300$ close galaxy pairs and an approximately similar number of isolated galaxies show that galaxy pairs tend to exhibit slightly but significantly higher levels of asymmetry. For consistency with these authors, we repeated our statistical analysis using the k-sample Anderson–Darling test \citep{Scholz1987}, but the results were unchanged.  However, it is not entirely trivial to compare our results with those in the literature. Adopting $A_F > 1.26$ as the criterion for asymmetry, the asymmetry fraction in our study is $5\pm 2\%$ for non-mergers and $13\pm 2\%$ for mergers. The uncertainty is determined by resampling the $A_F$ values 100 times according to their uncertainties, which consist of both statistical and systematic uncertainties, as explained in \citet{Yu2020ApJ...898..102Y}. However, the asymmetry fraction of non-mergers and mergers becomes $12\pm 2\%$ and $14\pm 2\%$, respectively, after we matched the samples one-to-one. The difference is no longer significant. Using a similar criterion, previous studies have reported a wide range of asymmetry fractions, from 9\% to 22\% for isolated galaxies (e.g., \citealt{Haynes1998AJ....115...62H,Matthews1998AJ....116.1169M,Espada2011A&A...532A.117E,Bok2019MNRAS.484..582B}, to $16\%-26\%$ for galaxies in clusters \citep{Scott2018MNRAS.475.4648S}, and 27\% for galaxy pairs \citep{Bok2019MNRAS.484..582B}.  It is important to note that our profile asymmetry measurements have been corrected for the effects of S/N, which can be important \citep{Watts2020MNRAS.492.3672W}, but not all authors follow this practice. Sample differences in stellar mass poses a further complication. While the dependence of profile asymmetry on stellar mass is controversial (e.g., \citealt{Espada2011A&A...532A.117E,Reynolds2020MNRAS.499.3233R,Watts2020MNRAS.492.3672W,Manuwal2021arXiv210911214M}), N. Yu et al.\ (in preparation) find that massive galaxies tend to have more symmetric \HI\ profiles. Our present GOALS sample is biased toward high-mass galaxies, and the control sample was carefully chosen to match them in terms of stellar mass, both spanning $M_* \approx 10^{10.25}- 10^{11.25}\,M_\odot$. By contrast, the sample of \citet{Bok2019MNRAS.484..582B} covers $M_* \approx 10^6-10^{11}\,M_\odot$.

\begin{figure}
\includegraphics[height=0.30\textheight]{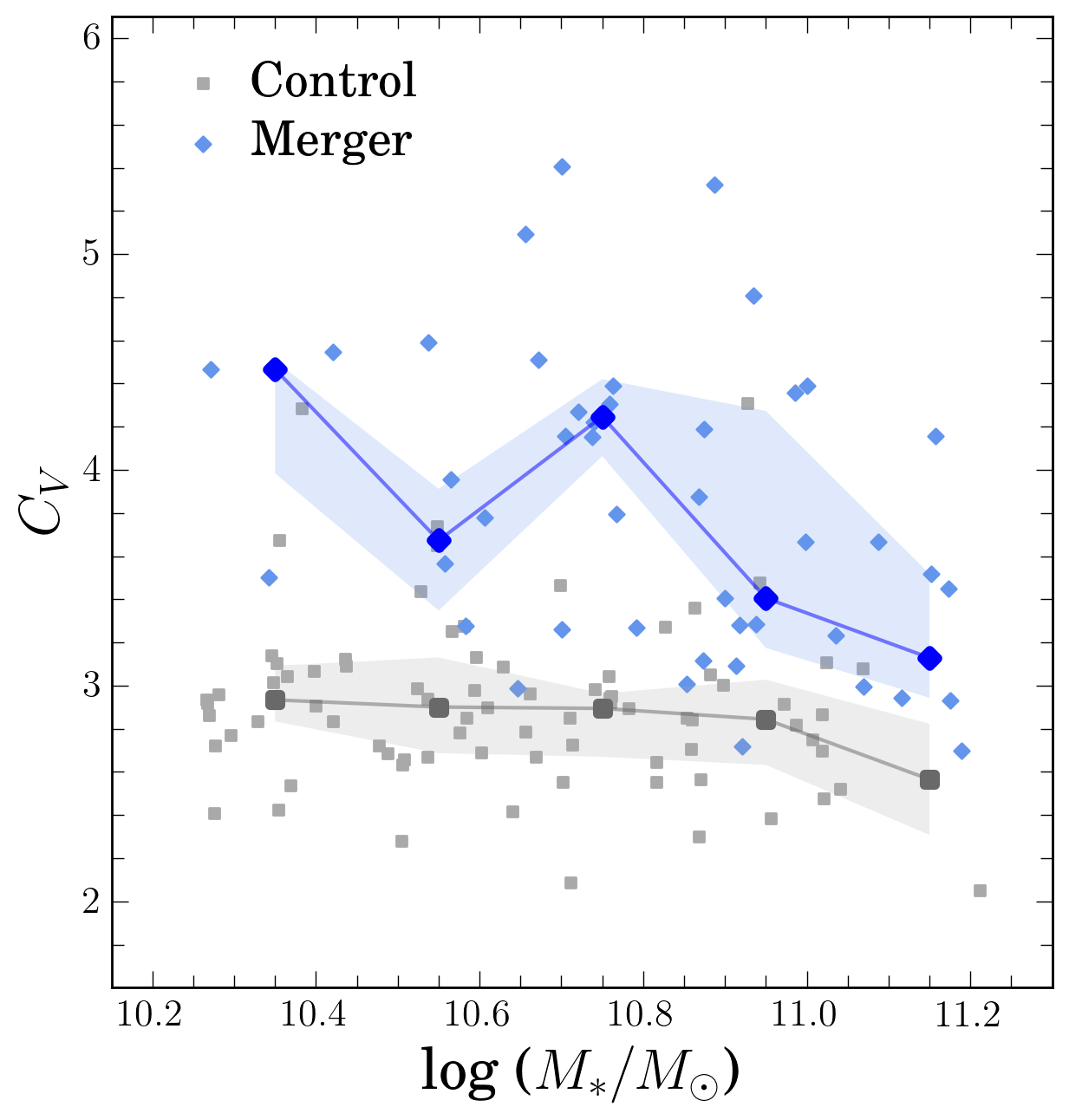}
\caption{The variation of profile concentration ($C_V$) with stellar mass ($M_*$) of the merger (blue diamond) and control (grey square) samples. The large symbols and the corresponding connecting lines give the median values of the objects in bins of 0.2 dex in $\log (M_*/M_\odot)$ from 10.25 to 11.25, and the shaded regions are the 25th and 75th percentiles of the distribution.}
\label{fig:CV_M}
\end{figure}

Whereas neither the width nor the asymmetry of the \HI\ line provides a clear signature of mergers, the detailed {\it shape}\ of the line profile does.  As illustrated for some examples in Figure~\ref{fig:opt-hi}, mergers typically display single-peaked \HI\ profiles, in strong contrast to the double-horned profiles characteristic of their non-merger counterparts.  Yu et al.\ (2022) demonstrate that the parameter $C_V$ introduced in \cite{Yu2020ApJ...898..102Y} effectively captures the range of line profile shapes observed in galaxies.   After controlling for projection effects due to inclination angle, \HI\ profiles of normal galaxies depend systematically on galaxy mass, as a consequence of both their internal distribution of atomic gas and their velocity field: massive disk galaxies of earlier Hubble type tend to have double-horned \HI\ profiles (low $C_V$), while the \HI\ spectra of low-mass, late-type disk galaxies are preferentially single-peaked (high $C_V$).  This trend is evident even within the relatively limited range of stellar mass covered by our control sample (Figure~\ref{fig:CV_M}).  It can be clearly seen that, at a given stellar mass, mergers have significantly larger values of $C_V$ than non-mergers, in support of the visual impression conveyed by Figure~\ref{fig:opt-hi} that the \HI\ profiles of mergers tend to be single-peaked, despite the fact that all of the GOALS objects are massive galaxies and therefore {\it should}\ be double-peaked.  Within our sample, 67\% of the mergers have single-peaked or flat-topped profiles ($C_V>3.4$), to be compared with only 12\% for the control sample.  The violin density plot in Figure~\ref{fig:vln-dis}e further reinforces the conclusion that the $C_V$ values of the merger and non-merger samples are significantly different, as does the K-S test, which yields $p < 10^{-4}$ (Table~\ref{tbl:ks-rst}).  

\begin{deluxetable}{cc} %[hbt!]
\small\addtolength{\tabcolsep}{25pt}
\tabletypesize{\normalsize}
\tablenum{5}
\tablecaption{Comparing Merger and Non-merger Galaxies}
\tablehead{
\colhead{Parameter} & \colhead{K-S Probability $p$} \\
\colhead{(1)} & \colhead{(2)} }
\startdata
$M_*$ & $0.60^{+0.1}_{-0.1}$\\
$V_\mathrm{85}$ & $0.36^{+0.02}_{-0.01}$ \\
$A_F$ & $0.40^{+0.02}_{-0.02}$ \\
$A_C$ & $0.48^{+0.01}_{-0.02}$ \\
$C_V$ & $< 10^{-4}$ \\
$\Delta V$ & $0.033^{+0.004}_{-0.003}$ 
%\fHI\ & $0.35^{+0.01}_{-0.01}$
\enddata
\tablecomments{
Column (1): Properties of the galaxy or its \HI\ profile.
Column (2): Median $p$ of the K-S test comparison between merger and non-merger (control) galaxies, based on 1000 bootstrap realizations of the two samples.  Error bars are derived from the 5 to 95 percentile range.
}
\label{tbl:ks-rst}
\end{deluxetable}

Lastly, we examine $\Delta V$, the relative offset between the radial velocity as derived from the \HI\ line and the systemic velocity of the galaxy, which we assume to be well-represented by the optical velocity measured through nebular emission lines from ionized gas or from stellar absorption lines.  Although the kinematics of the ionized gas may differ from that of the stars in galaxy mergers, previous studies have not found significant systematic difference between them (e.g., \citealt{Medling2014ApJ...784...70M,Barrera-Ballesteros2015A&A...582A..21B}).  The distribution of $\Delta V$ of mergers differs significantly from that of non-mergers ($p = 0.033^{+0.004}_{-0.003}$; Table~\ref{tbl:ks-rst}).  In particular, mergers have a prominent tail of positive values in $\Delta V$ (Figure~\ref{fig:vln-dis}f).

\section{Discussion and Summary} \label{sec:summary}

Major mergers, particularly those involving gas-rich galaxies, constitute important episodes in the lifecycle of massive galaxies.  Tides and shocks exert especially dramatic effects on the cold interstellar medium of the merging galaxies (e.g., \citealt{Barnes1996ApJ...471..115B,Barnes2002MNRAS.333..481B}), both on the molecular and atomic phases. Interferometric observations show that the tidal tails of galaxy mergers are \HI-rich and the \HI\ extends to tens of kpc \citep{Hibbard1996AJ....111..655H, Hibbard1999AJ....118..162H, Manthey2008A&A...484..693M}. The molecular fraction of galaxy mergers is higher than that in non-mergers, probably because \HI\ flows into the galactic center and then converts into molecular gas \citep{Larson2016ApJ...825..128L}. Simulations also show that the gas fraction in the inner region increases during the galaxy merger \citep[e.g.,][]{Blumenthal2018MNRAS.479.3952B}.
Mergers perturb not only the spatial distribution but also the kinematics of the atomic gas \citep{Horellou2001A&A...376..837H, Barnes2002MNRAS.333..481B, Struve2010A&A...515A..67S}.
Within this backdrop, \HI\ observations in principle can provide a sensitive tool to identify major mergers within the general galaxy population, complimentary to and perhaps even superior than traditional methods based on optical imaging \citep[e.g.,][]{Conselice2014ARA&A..52..291C}.  
However, major mergers comprise only a minority of the galaxy population (\citealt{ Lopez-Sanjuan2009AA...501..505L,Lopez-Sanjuan2009ApJ...694..643L, Lotz2011ApJ...742..103L, Casteels2014MNRAS.445.1157C, Ventou2017A&A...608A...9V}), and securing spatially resolved \HI\ maps of large, unbiased samples of galaxies beyond the local Universe are impractical for the near future. We are motivated to explore whether mergers imprint recognizable features on their integrated \HI\ line profile.  If so, then large extragalactic single-dish \HI\ surveys can be exploited to identify merger candidates efficiently.

We investigate this problem by analyzing the integrated \HI\ spectra of a subset of the galaxies in GOALS, a well-studied sample of nearby, bright starburst galaxies that encompasses a wide range of gas-rich systems along the merger sequence. With the aid of new observations acquired using FAST and existing data collected from the literature, we assembled integrated \HI\ spectra for 45 mergers and a control sample of 80 non-mergers matched in stellar mass from $M_* = 10^{10.25}$ to $10^{11.25}\,M_\odot$.  We analyzed the \HI\ spectra using the recently developed ``curve-of-growth'' method of \cite{Yu2020ApJ...898..102Y} to measure a number of line profile parameters that may be potentially useful for distinguishing between 43 merging and 43 non-merger galaxies with matched stellar mass. 

Intriguingly, we find that galaxies experiencing mergers do not exhibit notably broader lines compared to their non-merger counterparts.  Part of the difficulty stems from our inability to correct the observed line widths for projection effects, given the lack of a clear plane of symmetry for ongoing mergers, which is further compounded by the sizable intrinsic scatter in the Tully-Fisher relation \citep{Tully1977A&A....54..661T} for galaxies of earlier Hubble type (e.g., \citealt{Neistein1999}; \citealt{Ho2007}; \citealt{Williams2010}).  Contrary to the conclusion of \citet{Bok2019MNRAS.484..582B}, the mergers in our study do not show any obvious excess in asymmetry. This inconsistency is not unexpected, because \citet{Bok2019MNRAS.484..582B} selected their control sample using very strict criteria for isolation, while our control sample consists of more representative non-merger galaxies. While our results may suffer from small-number statistics---the sample of galaxy pairs in \citet{Bok2019MNRAS.484..582B} is significantly larger than ours---it is crucial to note that the extended \HI\ disk of galaxies is inherently fragile and easily susceptible to a variety of external perturbations.  Asymmetry in the gas distribution can be induced by gas accretion from the large-scale environment (e.g., \citealt{Bournaud2005A&A...438..507B,Sancisi2008AARv,Lagos2018MNRAS}), minor mergers of satellites (e.g., \citealt{Zaritsky1997ApJ}), flyby interactions (e.g., \citealt{Mapelli2008MNRAS}), and ram pressure stripping (e.g., \citealt{Gunn1972ApJ,Kenney2004AJ}).  Ordinary field galaxies have higher \HI\ asymmetry compared to truly isolated galaxies \citep{Espada2011A&A...532A.117E}, and even some of the isolated galaxies studied by \cite{Bok2019MNRAS.484..582B} are highly asymmetric.  Our control sample consists of galaxies that inhabit a wide range of environments \citep{Yu2020ApJ...898..102Y}, and hence its \HI\ asymmetry distribution is correspondingly diverse (Figures~\ref{fig:vln-dis}c and \ref{fig:vln-dis}d). We are forced to conclude that neither the width nor the asymmetry of the \HI\ profile offers an effective diagnostic to identify mergers.
%
% More unexpectedly, and contrary to the conclusions of \citet{Bok2019MNRAS.484..582B}, the mergers in our study do not show any obvious excess in line asymmetry. While our results may suffer from small-number statistics---the \citet{Bok2019MNRAS.484..582B} sample is significantly larger than ours---it is crucial to note that the extended \HI\ disk of galaxies is inherently fragile and easily susceptible to a variety of external perturbations.  Asymmetry in the gas distribution can be induced by gas accretion from the large-scale environment (e.g., \citealt{Bournaud2005A&A...438..507B,Sancisi2008AARv,Lagos2018MNRAS}), minor mergers of satellites (e.g., \citealt{Zaritsky1997ApJ}), flyby interactions (e.g., \citealt{Mapelli2008MNRAS}), and ram pressure stripping (e.g., \citealt{Gunn1972ApJ,Kenney2004AJ}).  Ordinary field galaxies have higher \HI\ asymmetry compared to truly isolated galaxies \citep{Espada2011A&A...532A.117E}, and even some of the isolated galaxies studied by \cite{Bok2019MNRAS.484..582B} are highly asymmetric.  Our control sample includes galaxies in different environments, from voids to clusters \citep{Yu2020ApJ...898..102Y}, and its distribution of \HI\ asymmetry is broad and complex (Figures~\ref{fig:vln-dis}c and \ref{fig:vln-dis}d).  We are forced to conclude that neither the width nor the asymmetry of the \HI\ profile offers an effective diagnostic to identify mergers.

Much more promising is the line shape.  As described in Yu et al.\ (2022), the line concentration parameter $C_V$, defined simply as the ratio of the velocity widths enclosing 25\% and 85\% of the total flux (Figure~\ref{fig:cv-trend}), conveniently describes line profiles that vary from double-horned to single-peaked. According to the statistical analysis of Yu et al. (2022), at a given inclination angle, low-mass, dwarf galaxies tend to be single-peaked (large $C_V$), and high-mass galaxies of earlier type (having larger bulges) are preferentially double-horned (small $C_V$).  Figures~\ref{fig:opt-hi} and \ref{fig:CV_M} clearly demonstrate that at a fixed stellar mass, mergers differ markedly from non-mergers: in view of their large stellar masses, mergers should possess double-horned \HI\ profiles, but, instead, they predominantly show single-peaked profiles. We suggest that the $C_V-M_*$ relation (Figure~\ref{fig:CV_M}) can be used as an effective tool to select massive galaxies that have a high probability of being mergers.

The $C_V-M_*$ relation has a simple physical underpinning. The integrated \HI\ profile reflects the convolution of the spatial distribution of the atomic hydrogen and the velocity field of the gas, which is governed by the gravitational potential of the galaxy. Late-type, low-mass galaxies have intrinsically single-peaked \HI\ profiles because the baryons, including the \HI, resides in a small inner radial range of the dark matter halo, such that the rotation curve keeps rising in the \HI\ disk \citep{Oh2015AJ....149..180O}.  
Massive, gas-rich spirals have double-horned \HI\ profiles because their neutral atomic hydrogen distribution has a central deficit \citep{Swaters2002A&A...390..829S} 
and because they have a steeply rising and then flattened rotation curve on account of their prominent bulge and less dominating dark matter halo \citep{de2008AJ....136.2648D}.  Mergers, despite being massive systems, systematically depart from this expectation presumably because their \HI\ gas is strongly centrally concentrated.  They have no central \HI\ hole.  The gravitational torques from the merger drive gas inflow, as observed (e.g., \citealt{Hibbard1994AJ....107...67H,Iono2004ApJ...616..199I}) and predicted by numerical simulations (e.g., \citealt{Barnes1996ApJ...471..115B,Barnes2002MNRAS.333..481B}).  The merger process naturally induces highly perturbed kinematics on the gas, as observed in the enhanced relative velocity offsets of the \HI\ ($\Delta V$).

\begin{acknowledgements}
We thank an anonymous referee for very helpful comments and suggestions.  This work was supported by the National Science Foundation of China (11721303, 11991052, 11903003, 12073002), China Manned Space Project (CMS-CSST-2021-A04, CMS-CSST-2021-B02), and the National Key R\&D Program of China (2016YFA0400702). We used data from FAST (Five-hundred-meter Aperture Spherical radio Telescope), a Chinese national mega-science facility operated by National Astronomical Observatories, Chinese Academy of Sciences. We thank Lister Staveley-Smith, Bi-Qing For, Ningyu Tang, Hongwei Xi, Barbara Catinella, and Xuanyi Lyu for useful advice and discussions. This research used the NASA/IPAC Extragalactic Database, which is funded by the National Aeronautics and Space Administration and operated by the California Institute of Technology. 
\end{acknowledgements}



\end{document}